\documentclass[preprint]{aastex}

\shorttitle{Vertical Disturbances of the ISM}
\shortauthors{Walters and Cox}

\begin{document}

\title{Models of Vertical Disturbances in the Interstellar Medium}
\author{Michael A. Walters \and Donald P. Cox}
\affil{Department of Physics, 1150 University Ave., University of
Wisconsin--Madison, Madison, WI 53706}
\email{walters@wisp5.physics.wisc.edu, cox@wisp.physics.wisc.edu}

\begin{abstract}
This paper describes some interesting properties of waves in, and
oscillations of, the interstellar medium in the direction normal to the 
plane of the Galaxy. Our purpose is to examine possible reasons for four 
observed phenomena: the falling sky in the northern hemisphere; the apparent 
presence of clouds in absorption spectra when a sightline is occupied 
primarily only by warm intercloud gas; the peculiar structuring of spiral 
arms involving clumps, spurs, and feathering; and the existence of an 
abundance of high stage ions far off the plane of the Galaxy.

We explored the reaction of the interstellar medium - in the vertical
direction only - to large imposed disturbances (initial displacements,
expansive velocities, and compressions), and to the introduction of small
amplitude waves via oscillation of the midplane. Our findings included: 1)
the anticipated growth in amplitude of high frequency waves with height; 2) the four
lowest normal modes for the oscillation of the atmosphere as a whole, as
functions of the height of the outer boundary; 3) the time for material to
`bounce' from one unusually dense state to the next as a function of height;
and 4) the tendency for the disk to develop a hot outer halo, either after
the passage of a single shock from a large event, or in response to a
continuous stream of small amplitude waves.

We discovered that three of the four observed phenomena targeted are likely 
to be closely connected. Following a large expansion, material 
near the plane falls back
first, with material at higher $z$ then falling in upon it. This provides
precisely the sort of velocity segregation observed in the northern sky, at
about 50 Myr after the event.  In addition, this bounce time (and/or the period
of the subsequent smaller oscillations, which is about twice the bounce
time) may be linked to structure in the spiral arms, with vertical
oscillations having been provoked by initial compressions in the arms. Oscillations of 
the fundamental symmetric (breathing) mode of the ISM also produce
substantial disturbances in the outer atmosphere. This can result in the 
production of an extensive layer of hot gas overlying the cooler disk material,
i.e. a hot galactic halo with a significant population of high stage ions. 
Hence, three of the four phenomena may be natural 
results of the simple existence of strong local compressions at the spiral 
arms, and the associated vertical motions in a thick galactic disk.  Finally, the 
somewhat mysterious appearance of clouds 
in some absorption spectra can be produced by small amplitude 
waves in the ISM. Under the
right conditions, clouds will seem to appear through ``velocity crowding,'' 
when in fact there are no density concentrations in space.

\end{abstract}


\section{INTRODUCTION}

In this paper, we discuss the results of a range of experiments performed
with a 1-dimensional hydrocode to determine the characteristics of 
vertical oscillations of the interstellar medium. We were interested in the 
dynamics of the ISM in the direction normal to the plane of the Galaxy (i.e. 
the vertical direction).  Because our model treats the distribution of the ISM 
as plane parallel, we sometimes refer to this as the ``galactic 
atmosphere problem.'' Our experiments were aimed toward shedding some light on 
four observed aspects of the interstellar medium.

The first of these aspects is the fact that in roughly half of the northern
sky, 21 cm observations indicate that there are two primary components to
the material distribution, a low velocity component and another, comparable
in column density, which is falling toward the plane with a velocity on the
order of 50 km s$^{-1}$ (e.g. Stark et al. 1992; see Figure \ref{fig20} for a 
typical profile). This infalling component has been studied
extensively (e.g. Danly et al. 1992) and named the IV (Intermediate Velocity)
Arch. Our experiments
explored dynamical situations that could lead to such velocity segregation,
a circumstance that we refer to loosely as the ``falling sky.''

The second aspect is revealed by the UV absorption line spectrum of HD 93521
(Spitzer \& Fitzpatrick 1993), a star which has a $z$ distance of about 1.5
kpc and lies behind the IV Arch. The H~I spectrum in that direction shows the
usual two components, with considerably more than thermal broadening, and
one rather narrower component at roughly -10 km s$^{-1}$\ heliocentric
velocity. On the other hand, the UV spectrum (with the reduced thermal
broadening of the lines of the heavier elements) shows a total of about 10
components, which have been interpreted as localized regions of enhanced
density. However, it is well known that velocity crowding (caustics in the
velocity distribution) can produce spectral features in an otherwise uniform
medium. Hence we performed experiments to explore whether the features
observed toward this star might in part derive from typical interstellar
velocity irregularities rather than density concentrations in space.

The third aspect involves the structure of spiral arms in galaxies. Such
arms are generally not entirely smooth and continuous, having localized
regions of enhanced contrast, in a pattern that could be called
feathering. One gets the impression that (as in the wake of a boat) there
are wavelets that are not traveling normal to the locus of the arms. In
addition, there are larger irregularities - pieces of arms or spurs - that
branch off of the smooth locus of the more regular pattern. Martos \& Cox
(1998) began to explore the vertical structure anticipated for a spiral arm,
including the thick disk and the significant nonthermal pressure. They used
only a two-dimensional model, but speculated on some of the potential 
three-dimensional characteristics, including the possibility that arm substructure
might be related to resonances in the induced vertical motions. In this
paper we have performed two sets of experiments useful in exploring the
relevance of this idea, determining the normal modes of vertical
oscillations and the timescale for interstellar material to return to the
galactic plane after being set into motion away from it.

The fourth aspect is related to the substantial quantities of high stage
ions (in particular C~IV, N~V, and O~VI) found well off the plane of the
Galaxy, with apparent scale heights up to several kpc (Savage \&
Massa 1987). A variety of possible sources for these ions have been
suggested. Here we are exploring only one, a possibility related to the
suggestion by Sciama (1972) that the outer boundary of the cool interstellar
medium might be chromospheric, with a layer of hot gas beyond that. Our
final set of experiments was aimed at examining the structure of the outer
edge of the disk resulting from mechanical disturbances within it, to learn
whether the transition to a hot outer layer is a natural consequence.

\section{THE CODE}

The simulations in this paper were performed with a 1-D plane parallel
Lagrangian hydrocode. Unlike some other Lagrangian codes (e.g. Bowers \&
Wilson 1991), ours did not make use of a mass coordinate for simplifying the
computation of derivatives. Instead, our code was constructed to maintain
second order accuracy by differentiating (and interpolating) explicitly in
space with piece-wise parabolic fitting (Walters 1993). This method allowed
us to maintain good resolution in all parts of the grid by using zones of
initially uniform spacing. It is well suited for cases in which the grid
spans several orders of magnitude in density, such as the galactic
atmosphere problem investigated in this paper. The code was otherwise
similar to several others constructed in our group for the study of
supernova remnants (e.g. Slavin \& Cox 1992; Cui \& Cox 1992), based on
the methods of Richtmyer \& Morton (1967).

Several additional features of the code are worth mentioning. Although this
was a one fluid model, it allowed for the possibility of different electron
and ion temperatures via the methods described in Cui \& Cox (1992) and
Cox \& Anderson (1982). It followed the ionization of hydrogen and helium
explicitly, including collisional ionization, radiative recombination, and
an additional low level of ionization (and heating) nominally due to cosmic
rays. The code also followed the heating and cooling rates for each parcel
(with heating primarily due to compressions and shock waves). For hydrogen,
cooling terms from bremsstrahlung, radiative recombination, and collisional
excitation were computed explicitly. All other elements were assumed to
radiate at their collisional equilibrium rates, using a rate table produced
from the Raymond and Smith code (Raymond, Cox, \& Smith 1976) with
standard cosmic abundances. In order to avoid excessively low temperatures,
all cooling except that due to recombination of hydrogen was turned off
below 10$^4$\ K. Electron thermal conduction was also included in the code. 
Most of these features had little effect on the overall behavior of the 
atmosphere, but allowed for a more realistic study of the outer layers of 
the galactic disk.

\section{THE HYDROSTATIC MODEL}
\label{hydrostatic}

Our model for the vertical distributions of density and pressure was similar
to that of Model A of Martos \& Cox (1998), based on the hydrostatic
solution of Boulares \& Cox (1990). The density distribution consisted of
the smoothed contributions from molecular and cold atomic clouds as well as
the warm neutral and the warm ionized components. Each component had its own
total column density, scale height, and form of distribution. The total
number density of nuclei was given by:

\begin{eqnarray}
n_o\left( z\right) &=&\left\{ 0.6\exp \left[ -\frac{z^2}{2\left( 70\textrm{ pc
}\right) ^2}\right] +0.37\exp \left[ -\frac{z^2}{2\left( 135\textrm{ pc}
\right) ^2}\right] \right\} \textrm{ cm}^{-3} \nonumber \\
&+&\left\{ 0.1\exp \left( -\frac{\left| z\right| }{400\textrm{ pc}}\right)
+0.03\exp \left( -\frac{\left| z\right| }{900\textrm{ pc}}\right) \right\} 
\textrm{ cm}^{-3} .
\label{eqn1}
\end{eqnarray}

The gas consisted of hydrogen and helium atoms, in the ratio of 10:1. (The 
presence of other elements was neglected when computing the bulk properties of 
the gas such as density and pressure.) The
electron density was inferred from the density of nuclei and the ionization
level. In the initialization, a constant cosmic ray ionization rate was
included which was sufficient to maintain 25\% ionization of hydrogen at a $z$
distance of 5 kpc from the galactic plane (assuming the equilibrium density
at that height).

We used the gravity given by Bienaym\'{e}, Robin, \& Cr\'{e}z\'{e} (1987),
as fit by Martos \& Cox (1998):

\begin{equation}
K\left( z\right) =8.0\times 10^{-9}\left[ 1-0.52\exp \left( -\frac{\left|
z\right| }{325\textrm{ pc}}\right) -0.48\exp \left( -\frac{\left| z\right| }{
900\textrm{ pc}}\right) \right] \textrm{ cm s}^{-2} .
\label{eqn2}
\end{equation}

Given the density and gravity distributions, one can derive the total
pressure as a function of height, the result depending somewhat on the
assumptions about the outer boundary. Our assumption was that the density
followed the form described above until the outer boundary was reached, at
which point the pressure was taken as purely thermal and then integrated
downward. Beyond the boundary, the density and pressure were taken to be
zero, making the outer boundary a free surface. The location of the boundary
was most often 5 kpc, but that height was varied somewhat for some of the
experiments described later. (Note: Having the pressure drop rapidly to
thermal near the outer boundary while the density remained finite had the
effect of terminating the disk with a layer of low signal speed. However,
this oddity at very high $z$ appears to have had little effect on the 
results of our experiments.)

The pressure forms assumed for the calculation were thermal pressure and a
pseudo-magnetic pressure. The thermal pressure was intended to include the
internal dynamical pressures of the denser components, but not the
large-scale motions actually being modeled. This was accomplished by
assuming a constant temperature of 10$^4$\ K for all components of the
static solution. There was no explicit inclusion of cosmic ray
pressure, known to be a very important component (Boulares \& Cox 1990). 
Instead, the entire nonthermal pressure was represented by the
pseudo-magnetic term.

Given the total pressure from the weight integration and the thermal
pressure from the density, ionization, and temperature, the magnetic
pressure was found by simple subtraction. Our form for the equilibrium
thermal pressure was:
\begin{equation}
P_{To}\left( z\right) =\chi\left( z\right) n_o\left( z\right)kT ,
\label{eqn3}
\end{equation}
where $\chi\left( z\right)$\ is the ratio of the total number of particles 
to the number of atoms or nuclei. It then followed that the equilibrium 
magnetic field would be given by: 
\begin{equation}
B_o(z)=\sqrt{8\pi P_{Bo}}=\sqrt{8\pi \left[ P_o(z) - \chi\left( z\right) 
n_o\left( z\right)kT\right]} ,
\label{eqn4}
\end{equation}
where $P_o(z)$ is the total initial pressure (thermal plus magnetic). When 
evolving the model over time, it was assumed that the magnetic field
was parallel to the galactic plane and the flux was frozen in the flow, so
that: 
\begin{equation}
B(z)=B_o\left( z_o\right) \frac{n(z)}{n_o\left( z_o\right) } .  
\label{eqn5}
\end{equation}
In this equation, the functions $B_o\left( z_o\right) $\ and $n_o\left(
z_o\right)$\ are the hydrostatic values for the Lagrangian parcel initially
located at $z_o$, and now found at $z$.

We employed both one-sided and two-sided models of the atmosphere. The
one-sided model was used for studying symmetric modes and wave propagation
into the atmosphere. In this model, the lower boundary represented the
galactic midplane and was rigid, meaning that there was no inflow or outflow
of gas. The
two-sided model was used to study asymmetric modes of oscillation. Figure 
\ref{fig1} shows the equilibrium pressure, density, magnetic
field, and ionization distributions of the two-sided hydrostatic model out
to $\pm$5 kpc. The one-sided atmosphere was the same except that it was
limited to that shown in the right half of each graph ($z \geq 0$).

\section{TESTING THE PROGRAM}
\label{secSmallOsc}

We tested the hydrocode with
several problems that have analytical solutions. One such problem was that
of a piston moving at a constant speed down a 1-D tube containing a uniform
medium. In this case, we found that the speed of the shock produced by the
piston was in excellent agreement with theory, with
differences usually less than 1\%. 
Our results were equally good when magnetic pressure was included in the 
calculation.

Another interesting test was to observe the growth of small waves as they
propagated upward in our model atmosphere. Using a one-sided disk, we
oscillated the midplane boundary up and down sinusoidally with a very small
amplitude. The amplitudes and positions of the extrema in velocity were
tracked as they propagated, and the results were compared with the WKB
approximation, which tells us that $\rho _ocA^2$\ is constant vs. $z$ (see
Appendix), where $\rho _o$\ is the undisturbed density distribution, $
c^2=\partial P/\partial \rho $, and $A$\ is the amplitude. In our case, the
signal speed is given by $c^2=(P_{To}+2P_{Bo})/\rho _o$. Again, we found good
agreement between theory and the results of our program.  For example, one test 
case had oscillations with a period of 5 Myr and an
amplitude of 0.01 km s$^{-1}$\ at the midplane.  At this period, the wavelengths are
roughly one third of a scale height, and the behavior should not necessarily
correspond precisely to the short wavelength limit. Nevertheless, we found that 
the growth in amplitude closely approximated the WKB result. (The amplitudes 
grew by factors of 10, 20 and about 60 at heights of 2, 3.6, and 5 kpc.) 
However, there was a small asymmetry between the positive and negative
velocities that appeared to decrease with the passage of successive waves.

\section{RESPONSE TO A LARGE DISTURBANCE}
\label{large}

In this section, we present the results of our investigation (via numerical
experiments) into the normal modes of vertical oscillation of our model ISM.
We attempted to find these modes, along with the corresponding resonant 
frequencies, by introducing a large disturbance into our model ISM. We 
succeeded in discovering both antisymmetric modes (which we call sloshing 
modes, for reasons discussed later) and symmetric (breathing) modes. We found 
two distinct frequencies for each type, and also discovered how those
frequencies depend on the assumed height of our outer boundary. For
comparison, we undertook a separate linear normal modes analysis (see
Appendix) assuming a pressure node at the
top of the atmosphere, and found almost exactly the same frequencies as
reported below from the numerical experiments.

Our interest in these frequencies arises from the hypothesis that resonances
between the vertical modes and spiral density waves might lead to
reinforcement of particular patterns within the arms, as suggested in Martos
\& Cox (1998). This could produce substructure in the spiral arms, such as
feathering, spurs, or ``beads on a string.''

In addition to the periods of normal modes, we also examined the response of
the ISM to a large expansive event, provoked by an initial strong
compression of the central portion of the disk. We explored the time
required for the gas to bounce from one compressed state to the next as a
function of the strength of the event and of the distance from the midplane.
The motivation was similar, to understand this characteristic bounce time as
a possible indicator of the distance between compressed regions along the
length of a spiral arm, or between the arms themselves. It also gives
an idea of the recovery time of the disk after any major perturbation, such
as impact by a high velocity cloud.

\subsection{Density-Antisymmetric (Sloshing) Modes}

In our search for antisymmetric modes, the entire two-sided atmosphere was
initialized as if it were in hydrostatic equilibrium, then shifted in height by a
constant amount and released. For our first simulation, we chose an
offset of 100 pc and atmospheric boundaries at $\pm$3 kpc.

Our program keeps track of the properties of certain designated parcels,
chosen to show what different regions of the atmosphere are doing. As a
rule, the parcels examined have equilibrium heights (before the
displacement) of $z_o$\ = 0, 10\%, 20\%, 50\%, and 100\% of the upper boundary
height (i.e. the top of the atmosphere, $z_{\max}$). In the results below, we
show the locations of these parcels' upper boundaries versus time. We
also followed changes in the total energy in the grid. These results were then 
Fourier transformed to find the periods of oscillation.

Figure \ref{fig2} shows the history of the locations of these tracer parcels
over the course of a 3 billion year simulation. As the atmosphere oscillates
vertically, the motions of the individual parcels appear to be sinusoidal with
beats of roughly five oscillations each, implying the
superposition of two frequencies differing by about 20\%. Note also that the
modulation due to the beats is relatively small for the parcels at
intermediate distances from the midplane, implying that one of the
frequencies has a node in the displacement in that vicinity. This is just
what would be expected for a mixture of
the fundamental antisymmetric mode and its first overtone. The amplitude of
the modulation is quite large at the midplane and the upper boundary, requiring comparable wave
amplitudes there. The midplane is a displacement anti-node for both
frequencies. The outer boundary is a pressure node, corresponding to a
velocity and displacement anti-node. Also shown in Figure \ref{fig2} is the
history of the total energy (per unit area) in the grid. Note that this is the total energy
of \textit{both} sides of the atmosphere. The oscillations in this plot
arise from the weak heating term that acts when expansion cools the gas
below 10$^4$\ K. Because this happens with opposite phase on the two sides,
the oscillations in energy have twice the frequency of those in displacement.

We performed Fourier transforms on these histories, using the algorithm
given by Press et al. (1989). To determine the periods accurately, we
eliminated the early complications of the transient response by examining
only the last 2 billion years of the simulation. Figure \ref{fig3} shows the
power spectra resulting from the Fourier analysis. As expected, they are
dominated by two large peaks separated by about 20\% in frequency. 
The corresponding periods are 117.6 and 93.0 Myr, independent of height above the midplane. 
(For comparison, the periods from our linear normal mode analysis
are 118.5 and 93 Myr.)  Note that the relative amplitude of the higher frequency 
peak is low for tracer parcels at intermediate distances from the plane, in agreement 
with the expectation that it has a node in that vicinity.

Figure \ref{fig4} illustrates the structure in the atmosphere at 1.5, 2.0,
and 2.5 billion years. The ``sloshing'' back and forth (across the midplane) is 
evident, though complicated by the
contribution from the second mode that tends to move material at high $z$
opposite in direction from that at low $z$. This sloshing motion (which is why
we call this the sloshing mode) can be seen by observing changes in the pressure 
distribution between the three times. In a linear
approximation, the velocity distributions would be entirely symmetric, with
the velocity of the center occasionally opposite in direction from that at
great distances from the plane, depending on the amplitudes and phases of
the two modes. For the particular times shown, the velocities are everywhere
of the same sign (an accident of the phasing), but there is an obvious lack
of complete symmetry in the velocity field. This asymmetry could be due to 
any of several factors: the anticipated symmetry is with respect to the
unperturbed positions of the parcels; the flow includes the full
nonlinearities which should be most significant far from the plane where the
velocities are large; and the power spectrum shows very weak features at
other frequencies that may include contributions excited at the opposite
symmetry through round-off errors or asymmetries in the code. In spite of these
complications, however, the atmosphere oscillates at moderate amplitude at
just those frequencies expected from the linear analysis. In addition, the
oscillations persist for an extremely long time.

Repeating this simulation using atmospheres with various
maximum heights, $z_{\max }$, ranging from 1 to 5 kpc, 
we found that the periods increase
as the height increases, still in excellent agreement with the linear
analysis. The results are shown as the two solid lines in Figure \ref{fig5}.
Note that while the height of the atmosphere increases from 1 to 5 kpc, the
period of the fundamental mode goes from 104 to 133.3 Myr, an increase of only
30\%. In fact, our linear analysis tells us that the period should approach an 
asymptotic value as $z_{\max }$ increases, as the forces 
inversely proportional to the wavelength squared fade in importance relative 
to those inversely proportional to the scale height squared.

\subsection{Density-Symmetric (Breathing) Modes}

To excite modes that are symmetric about the galactic midplane, we
initialized the model with a symmetric compression of the lower atmosphere,
reminiscent of the compression observed in spiral arms. In the first
experiment, all parcels within 500 pc of the midplane in equilibrium were
compressed by a factor of 1.5, so that their heights decreased and their
densities increased by a factor of 1.5. Above (or below) this compressed
layer, all of the parcels retained their equilibrium properties, although they
were shifted downward (or upward) due to the smaller thickness of the 
compressed layer. For comparison with our previous simulation, we again 
chose an initial boundary height for the atmosphere of 3 kpc.

The results for a 3 billion year run are shown in Figure \ref{fig6}. Again
there is periodic motion with amplitude modulation. The Fourier transforms
of these histories are shown in Figure \ref{fig7}. Notice that in this case
not all parcels have the same oscillation frequencies. The motion of the
galactic midplane ($z_1$) corresponds to periods of 117.6 and 93.0 Myr, just
those of our previous asymmetric disturbance. However the rest of the
parcels and the energy show periods of 109.6 and 70.2
Myr (compared with 109.5 and 69.5 Myr for the two lowest symmetric modes
from the linear analysis).  And once again, the relative amplitude of the higher 
frequency component is low at intermediate heights, close to its node.  Ideally, 
one would expect the midplane to remain perfectly still in this experiment, but 
we find that a very slight amount of sloshing has been excited, which is 
probably due to round-off errors. These oscillations are much, much 
smaller than any of the other oscillations, and are only visible because the 
midplane is a velocity node for all symmetric modes. Any spurious sloshing mode 
frequencies are totally invisible (below noise level) in the oscillation spectra 
taken at other heights.

Figure \ref{fig8} shows the structure of the atmosphere at 1, 2, and 3 Gyr.
As expected the density, pressure, and temperature distributions are
symmetric about the midplane, while the velocity profiles are antisymmetric.
The velocity profiles at 1 and 3 Gyr show the
entire disk (both above and below the midplane) to be either expanding or
contracting, in other words ``breathing.''  However, at 2 Gyr the motion changes 
from expansion to contraction at about 2.3 kpc off the plane, revealing the 
presence of the overtone.

Repeating the simulation using several different boundary heights for the
atmosphere, as we did in the previous section, yielded the results shown as
dotted lines in Figure \ref{fig5}. Figure \ref{fig5} in fact illustrates all
of the anticipated characteristics of normal modes. The longest period
belongs to the sloshing mode with no displacement nodes. From there, the
periods decrease with increasing node number, alternating between symmetric
(breathing) and antisymmetric (sloshing) forms. The periods also increase 
gradually with increasing
atmospheric thickness, the specific rate depending on the detailed structure
of the atmosphere and in particular the $z$ dependences of gravity, density,
and the signal speed. As $z_{\max}$ goes to infinity, these periods should 
approach asymptotic values.

\subsection{Mixing Modes}

A final experiment was performed to excite both sloshing and breathing modes
with comparable amplitudes for an atmosphere with boundaries at $\pm$2.5
kpc. In this setup, a one-sided expansion was initialized by imposing a
uniform velocity gradient equal to 25 km s$^{-1}$\ kpc$^{-1}$\ on the upper
half of the atmosphere ($z \geq 0$) and leaving the other half (initially) at rest. The
resulting oscillations of the tracer parcels are shown in Figure \ref{fig9}. 
In the Fourier analysis of these motions, shown in Figure \ref{fig10},
this modulation pattern is found to correspond to periods of 114.3 and 102.6
Myr. The 114.3 Myr mode is exactly that of the first sloshing mode for an
atmosphere of this height, while the shorter period is very close to the
103.9 Myr value for the first breathing mode. (The linear analysis yielded
114.7 and 103 Myr, respectively.) The midplane, which is a displacement node
for the breathing modes, shows no beats and its spectrum contains only the
longer period. The amplitude modulation shown in Figure \ref{fig9} increases
monotonically with height until it is 100\% at the outer boundary. This is
reflected in Figure \ref{fig10} as an increase in the higher frequency
component until the two are equal at the outer boundary.

Comparison of Figures \ref{fig2}, \ref{fig6}, and \ref{fig9} shows that our
experiments regularly produced large amplitude modulation at the outer
boundary. We do not know whether this is due to an accident of the
initializations, nonlinearities in the early more dramatic motions, or
a natural property of this atmosphere. Note also that all experiments led to
spectra dominated by only two frequencies. In the mixed mode experiment, we
did not obtain all four of the previous frequencies, only the two lowest
ones. This pair of curiosities may indicate something fundamental about how
the atmosphere responds to major disturbances.

\section{THE TRANSIENT RESPONSES AND RETURN TIMES}

In the previous section, we were only interested in the long-term behavior
of the atmosphere, disregarding the atmosphere's initial (transient)
response to disturbances. As indicated previously, the initial response to
the disturbance is also of interest and we examine that next.

In the experiments on symmetric modes, one of our simulations had a
compression factor of 1.5, a total atmosphere height of 5 kpc, and a run
time of 3 Gyr. Figure \ref{fig11} shows the tracking of our tracer parcels
for the first 1 Gyr of this simulation. When the parcels first rebound from
the compression imposed on them, they are propelled to relatively large
heights. They then fall back toward the midplane, slowing down only as they
come close to their equilibrium positions. Oscillations subsequent to this
first bounce are smaller and rapidly approach the periodic pattern discussed
previously. The parcels very close to the outer boundary require more than
one bounce to settle. The midplane oscillation amplitude should in principle
be zero, but in actuality has a small asymmetric component. All other 
regions seem to settle after only one return.

A more detailed view of these transients can be seen in Figure \ref{fig12},
which shows the first 250 Myr of this simulation. Note that for this
illustration, the simulation has been redone with the first tracer parcel
set at 5\% of the peak height of the atmosphere, rather than at the nearly
stationary midplane.

We repeated the experiment for a range of initial compression factors and
two different boundary heights (5 kpc and 3 kpc). Defining the bounce time,
$t_{bounce}$, for a parcel as the time from the beginning of the experiment 
to time of its next minimum in height, the results for the atmosphere with the 5 kpc boundary 
height are shown in Figure \ref{fig14}.

As we can see in Figure \ref{fig14}, the bounce times for the parcels of the
inner disk depend very little on compression factor, which is a measure of
the strength of the disturbance. There is a slight decrease in bounce times
with increasing compression factor, perhaps because these parcels return
more rapidly when they hit the upper atmosphere harder. Also, these times
are only weakly dependent on the equilibrium heights of the parcels, with a
slight linear gradient that arises from the flattening of gravity and the
increase in the amplitude of the disturbance with increasing $z$. Conversely,
parcels very high in the atmosphere have significantly longer bounce times,
which is partly due to the longer time it takes the initial disturbance to
reach them, partly to the nearly constant gravity there, and partly to the 
rapid growth in the amplitude of the
disturbance with $z$. They also take progressively longer to return with
increasing compression factor. 
Much the same behavior is found when the boundary height is reduced to 3 kpc, though
the initial decrease in bounce time with increasing compression factor for
parcels near the plane is more pronounced. 
The overall result for moderate strength disturbances and either boundary height is that
the bounce times near the plane are about 40 Myr, increasing slowly with 
parcel height by roughly 20 Myr kpc$^{-1}$. For weaker
disturbances, the bounce time is somewhat longer, closer to the half period of
the first symmetric mode, but increases more slowly with height. On the whole,
bounce times for material near the midplane
lie between 40 and 50 Myr, increasing to about 70 Myr for material 1.5 kpc
off the plane, and to much longer times for the outer boundary. A fundamental
consequence of these results is explored in the next section, but a
discussion of additional possible significance is deferred to the
conclusions.

\section{THE FALLING SKY}

As discussed in the introduction, Spitzer \& Fitzpatrick (1993) have
observed the UV absorption spectrum of the Galactic halo star HD 93521,
identifying at least nine separate velocity components that they interpret
as interstellar clouds. These components congregate in two velocity regions,
one approximately stationary and the other moving towards the midplane at
roughly 50 km s$^{-1}$. This general structuring is typical of roughly
half of the northern sky, which is covered by the IV Arch. The numerical
experiments in this section are designed to show how a large disturbance in
the galactic atmosphere can naturally create such velocity segregation, and
that small velocity perturbations can produce the multiple velocity
components seen without any actual clouds (via velocity crowding, i.e.
caustics in the velocity field).

\subsection{Velocity Segregation}
\label{secVelocity}

As explored in the previous section, material may move away from the plane
following a large disturbance within the disk, returning after a
characteristic bounce time. But, because that return time increases with
height, material nearer the plane falls back sooner and accumulates near the
midplane at low velocity, with material further up continuing to rain down
upon it. If the velocities involved are large, the infalling outer
atmosphere will be bounded on the bottom by a shock between it and the
already decelerated material close to the plane. Owing to the considerable
density gradient in the infalling material, the spectrum during such infall
will show two fairly sharp velocity components, one at low velocity due to the nearly
stationary material near the disk, and one at an ``intermediate'' negative
velocity due to the material just about to pass through the shock. (For smaller
disturbances with lower velocities and no shock, the second feature will appear
at the extremum velocity of the infall as it just begins to slow.) Thus, the
observed velocity segregation is a natural consequence of a large
disturbance and should be found at a time roughly equal to the return time
for material a few hundred parsecs off the plane, which from our previous
discussion is expected to be 50 to 60 Myr, depending on the thickness of the
disk.

In our experiment to explore this phenomenon, we initialized the program
with the two-sided equilibrium atmosphere. A velocity profile was then
imposed such that the whole atmosphere was expanding. The velocity
rose/dropped linearly from $v=0$\ at $z=0$\ to a maximum/minimum of $v=\pm
60 $\ km s$^{-1}$\ at $z=\pm 1$\ kpc. For $\left| z\right| \geq 1$\ kpc, the
velocity was constant at $v=\pm 60$\ km s$^{-1}$. The equilibrium height of
the atmosphere's boundary was 5 kpc.

Figure \ref{fig18} illustrates the evolution of one side of this symmetric
model, showing the density and velocity profiles at several times. Also
shown is the predicted differential column density vs. velocity (i.e. the
velocity spectrum) that would be seen in the northern sky by an observer at 
the midplane. For 
comparison with HD 93521, we integrate the column density out to 1.5 kpc, 
the approximate distance to that star. We also use a lower limit
of 235 pc for the integration to account for the presence of the local
interstellar cavity. This particular limit was chosen to make the total
column density equal to 1.25$\times $10$^{20}$\ cm$^{-2}$, at least
initially, which is the value observed in that direction.

The solid line in the left column of Figure \ref{fig18} illustrates the
initialization. The absence of the lowest velocities in the spectrum is due
to the lower limit of the integration (at 235 pc). The dotted line, 20 Myr
later, shows the substantial amount of expansion involved, and that a
significant amount of material close to the plane has reached zero
velocity. The dashed line at 40 Myr shows that the velocities have reversed
out to about 3 kpc. The spectrum at that time has a peak at the extremum of
negative velocity, and again has a hole at zero velocity due to the lower
integration limit. The right column of Figure \ref{fig18} shows times of 50,
55, and 60 Myr, during which a region of nearly zero velocity expands out to
about 1 kpc from the midplane. Outside this region there is substantial 
infall. As time
progresses, the velocity of infall increases, while the column density of
infalling material decreases. At the latest time shown, the structure is
forming a shock which appears as an abrupt change in the velocity and as a
step in the density profile. (Note also that the density profile interior to
the shock differs very little from the equilibrium distribution.) At later
times the shock strengthens and passes out through the atmosphere. This
shock substantially heats the tenuous outer parts of the atmosphere (a
subject we will return to in the next section) and drives it back into
expansion, leading to the subsequent smaller amplitude oscillations explored 
in the previous sections.

Figure \ref{fig19} provides an enlarged view of the velocity spectrum at 55
Myr. The solid line is the actual spectrum, with strong peaks near -3 and
-48 km s$^{-1}$; the dotted and dashed lines show convolutions with the
thermal profiles for oxygen and hydrogen, respectively. The synthetic
spectrum for hydrogen closely resembles the 21 cm observations for many
regions in the northern sky in the general direction of the IV Arch (Danly
et al. 1992), in particular that shown by Spitzer \& Fitzpatrick (1993) in
the direction of HD 93521, reprinted as Figure \ref{fig20}. 
The details of these spectral results depend somewhat on the initial form of
the disturbance. Roughly speaking, a steeper initial gradient promotes more
mass to higher velocity. This results in a larger column density in the
infalling component when it is centered at a given velocity, or a higher
infall velocity for a given column density in the infalling component. Our 
initialization was selected
to produce comparable column densities in the two components when the
infalling component had a speed of roughly 50 km s$^{-1}$.

\subsection{Adding Velocity Substructure}

We have shown that the bi-modal velocity distribution observed in the
northern sky can be produced by vertical motions in the Galaxy. Next, we 
show how features resembling clouds can be produced. In the previous
section, we initialized our model with a velocity profile which has a
uniform gradient near the midplane and is flat away from the midplane. But
the interstellar medium has a considerable amount of velocity structure on
scales of a few kilometers per second, which needs to be included for a more
realistic assessment of the anticipated spectrum. Rather than trying to
introduce these small-scale structures in the initialization, however, we
chose to produce them by mechanical disturbances at the midplane, which then
propagated into the outer disk as waves.

The experiment in this section had exactly the same initial configuration as
that in the previous section, with one important difference. To produce
small waves, we applied a small sinusoidal motion to the midplane boundary.
The period for these oscillations was 10 Myr and the velocity amplitude was
5 km s$^{-1}$. There were no waves present prior to the beginning of the
expansion. Although this method does not reproduce the randomness of
interstellar turbulence, it nevertheless provides disturbances of the about
the right amplitude and spacing. It also gives us a clear example to
illustrate our point about caustics.

Figure \ref{fig22} illustrates the evolution of this model. Comparing this
to Figure \ref{fig18}, one can see that the behavior of the atmosphere as a
whole is very similar to the case without waves. The waves are of such low
amplitude that there is scarcely any density or pressure variation visible,
but they can be seen as small wiggles in the velocity profile near the
midplane, moving outward with time (until they are swept back by the
infall). At late times, these variations also appear as small inclined steps
in the density plot. On the other hand, the effect on the velocity spectrum
is quite pronounced, even at early times. At later times we can discern at
least 5 separate peaks, as can be seen in the spectrum at 55 Myr shown in
Figure \ref{fig23}.

These additional components arise from the convolution of the monotonic
density profile with the steps introduced into the velocity structure by the
small amplitude waves. The number of such steps increases with the
oscillation frequency, but at a significantly higher frequency they would
smear together to return to a nearly smooth distribution. In this particular
experiment, only 5 oscillations have occurred by the time the infall has
reached the desired state, and they have not quite managed to propagate as
far as the velocity minimum. Had there been waves in the experiment prior to
the expansion, there might have been even more high velocity components.

Comparing the spectrum of Figure \ref{fig23} with the 
21 cm profile and UV components toward HD 93521, shown in Figure \ref{fig20}, there are
several qualitative similarities. Both show two large concentrations of mass
in velocity space: one near 0 km s$^{-1}$ and the other near -50 km s$^{-1}$. 
Furthermore, each of these broad features consists of multiple components.
The inclusion of thermal broadening for hydrogen (dashed line), however,
obliterates the small scale structure, making the spectrum into a 
pair of bell-shaped features. This is very much like the observed profile of
Figure \ref{fig20}, the only significant difference being that the latter
contains a single much cooler component with low broadening.

Thus the combination of the increasing bounce times with increasing $z$ and
the presence of small amplitude waves (or, more realistically, the observed 
random motions) in the interstellar medium naturally produces a spectrum with 
the form and complexity of that of HD 93521, but with no density maxima actually 
associated with the velocity features! Our model includes no clouds whatsoever. 
All it requires is that our region
of the Galaxy have undergone a substantial amount of vertical expansion
roughly 50 Myr ago in order to produce the downfalling IV Arch that is
presently observed over roughly half of the northern sky.

One caveat to our model is that the distribution of material in the IV Arch
is observed to have column density variations from direction to direction
over the sky, almost surely implying that there are density concentrations
within it. It would be difficult to imagine that its return to the plane of
the Galaxy would not have led to density structuring, even if it were not
already present. However, velocity structuring within the ISM is just
as real and should frequently contribute to features in the spectra in low
column density directions. The fact that one feature in the spectrum of HD
93521 is noticeably cooler than all the others suggests that it might be the
one real cloud seen in the spectrum, the others being due to velocity
crowding in the diffuse intercloud component.

It is worth mentioning that these results are not peculiar to a one-dimensional 
model. Martos has conducted similar eexperiments in two dimensions, and has 
achieved similar results (personal communication; see also Fig. 16 in Martos \& 
Cox 1993).

\section{MAKING A HOT HALO}

As described in Section \ref{hydrostatic}, our equilibrium atmosphere
terminates at its outer boundary with a sharp cutoff in density, with
material just interior to the boundary having a temperature of 10$^4$\ K.
This is an unrealistic termination but was meant to be one that included no
bias concerning what actually occurs there. Our intent was to investigate
what does happen there, keeping in mind the hypothesis of Sciama (1972). He
suggested that the galactic atmosphere should pass through a chromospheric
state far off the plane and have a high temperature corona beyond that. This
transition could be produced by a variety of influences including the
dissipation of mechanical energy arriving from below, the eventual
significance of cosmic ray heating at very low densities, heating due to
flare-like activity, or dissipation of infall energy.

In this final set of experiments, we have explored how outward fluxes of
mechanical energy (like those invoked in our earlier experiments)
restructure the outer parts of the atmosphere, moving it away from our
initially assumed equilibrium. The basic result is that when the disturbance
produces a strong shock, or a series of shocks that are not too far apart,
there is a sudden transition at some height to a much higher temperature
regime. When the disturbance leads to a only a single strong shock, its
passage generates a hot outer layer which then eventually cools back to the
initial equilibrium. However, the cooling time is extremely long owing to
the low densities, and is longer still if the temperature rises beyond the
peak in the cooling curve. For small periodic perturbations, a ``hot halo'' 
can be built up by the successive passage of many waves or shocks. Also,
there is a significant resonance at which the halo produced is hotter and
more distended than it is otherwise. The power required in the
waves to maintain an appreciable hot halo is extremely small.

As a first example, we return to the simulation from the previous section
that has an expanding atmosphere with a maximum velocity of 60 km s$^{-1}$.
It produces a large segregation in velocity after $\sim $ 50 Myr, with an
outward facing shock whose velocity jump approaches the original expansion
speed, 60 km s$^{-1}$. Following this simulation for a much longer time, we
see that the shock moves outward through the infalling material, remaining
radiative but generating long lived ionization until it reaches a region of
the atmosphere with sufficiently low density that the cooling time exceeds
the shock transit time through the remainder of the atmosphere. The material
heated during that final phase is the transient hot halo in this case.
Subsequently, the inner denser parts of the halo cool faster than the outer
regions. Both the amount of material remaining hot and the temperature of
the hot material gradually decrease over time. This description is
complicated slightly by the further reverberations of the atmosphere, which
send additional weaker shocks into the halo.

These effects are illustrated in Figure \ref{fig24}, which shows the
density, temperature, and ionization structure of the model at very late
times: 400 and 600 Myr after the start of the simulation. The sudden drop in density
and rise in temperature at height of 5 kpc (the nominal boundary height) are
apparent. The temperature in the halo is roughly 10$^5$\ K, decreasing with
time. Note, however, that the halo persists for hundreds of millions of
years. The density in the halo is roughly 10$^{-5}$\ cm$^{-3}$, at which the
anticipated cooling time is consistently expected to be around 500 Myr.

Next we examined the effect a series of small waves or perturbations, with
particular interest in what would happen at the resonant frequencies
found in Section~\ref{large}. In this set of experiments, we considered a
one-sided atmosphere with an outer boundary initially at 5 kpc. We forced
the midplane to oscillate with a velocity amplitude of 2 km s$^{-1}$ and
explored a range of oscillation periods between 100 to 150 Myr, which
includes the lowest frequency resonant modes (which have periods of 129 and
133 Myr for this atmosphere). The perturbations were made small enough 
so that an individual oscillation would not cause a
dramatic change in the atmosphere, but would still be  on the scale of the 
velocity dispersion 
of the interstellar medium. In each case, we followed the evolution for 3 Gyr, and
the structures discussed below are those at the end of that time.

Figure \ref{fig25} shows the final structure of the atmosphere for an
oscillation period of 100 Myr (relatively far from the resonances), as well
as the net change in energy in the atmosphere during the evolution. A hot 
ionized halo with a temperature rising to roughly 2.5$\times $10$^5$\ K has 
formed above the cool disk boundary.  The density in the halo 
drops from $10^{-4}$\ to $10^{-5}$\ cm$^{-3}$\ over the range 5.3 to 6.4 kpc.
Examining the structure at other times, we found that this hot layer remains
approximately constant in size, position, and temperature. The atmosphere
has therefore reached a new equilibrium state in response to the energy
input from the oscillations. This can also be seen by looking at the net total 
energy in the atmosphere vs. time, shown in the lower right graph in Figure 
\ref{fig25}. The large oscillations are just the insertion and removal of $PV$
work as the lower boundary pushes and pulls against the midplane pressure. 
If one looks past these oscillations for an overall 
trend, one can see that the average energy is nearly constant. 

For an oscillation period of 133 Myr, the first sloshing mode resonance 
found in Section~\ref{large}, we again find a hot halo, similar to
but slightly hotter than in the previous experiment. Interestingly, the 
envelope of the energy curve appears to oscillate in this case, suggesting 
that the
driving frequency may be beating against a nearby resonance.

Figure \ref{fig27} shows the results for a period of 129 Myr, the first
breathing mode resonance. They are dramatically different from the previous
cases. The halo has become very distended, extending from 4 to 20 kpc, and
is significantly hotter, with a temperature of 10$^6$ K. 
The lower right panel shows the time evolution of the net energy per
unit area, relative to the initial equilibrium state. 
In this case, the mean total energy has a
significant increase, rising by about 2$\times $10$^8$ erg cm$^{-2}$\ over
the 3 Gyr run. Interestingly, the average energy seems to level off after
1.5 Gyr, but then turns up again near the end.

The results for 133 and 129 Myr are remarkably
different for having been produced by only a relatively small shift in period. Both
cases used resonant frequencies for the unperturbed atmosphere, but those
frequencies may not correspond exactly to resonances in the new atmosphere
with its lowered boundary for the cool gas and its hot outer layer. In order
to explore the possibility that there might actually be a strong resonance
for hot gas generation, suggested by the great differences between these two
results, we explored a wide range of driving periods.

Figure \ref{fig28} illustrates the change in the mean energy column density
over 3 Gyr runs for a wide range of periods of oscillation of the midplane.
The mean energy is taken as the average of the maximum and minimum in one
cycle, and the change as the difference between the values for the first and
last full cycles in the simulation. The curve shows a strong resonance with
peak response at 127 Myr. (A quick investigation for
smaller periods revealed the presence of additional resonances in the 
neighborhoods of 60 and 80 Myr.)

The simulation using the peak period of 127 Myr is shown in
Figure \ref{fig29}. The hot halo is even more dramatic than for the 129 Myr
case. The temperature now jumps up sharply and continues to rise, reaching
2$\times$10$^6$ K. Because material formerly in the cool part of the
atmosphere has been promoted into the halo, supplying a density in excess 
of $10^{-5}$\ cm$^{-3}$\ out to 11 kpc, the boundary between the two regimes has
retreated downward slightly.  In this case, the average energy does not
level off, but continues to increase and even to increase in rate toward the
end of the run, possibly due to a decrease in the radiative losses of the
hot gas as it is heated further.

Let us now consider the power involved in generating this remarkable structure. 
The only input into the model comes from
the (sound) waves generated by the midplane, which acts like a piston. If
the midplane were a source of ordinary outward moving sound waves then the
average power per unit area, or intensity $I$, would be: 
\begin{equation}
I =\frac 12\rho v^2c
\approx 6\times 10^{-8}\textrm{ erg cm}^{-2}\textrm{ s}^{-1} .
\label{eqn6}
\end{equation}
We have used the initial midplane values for the density ($\rho$) and signal
speed ($c$). Note that this result does not depend on the period of the
piston motion, given the constant velocity amplitude. In equilibrium
there are nearly equal amounts of power propagating upward and downward, so
the above estimate is an extreme upper limit. In reality, the pressure and
velocity variations are nearly a quarter period out of phase, making their
product average to a much smaller value. The difference between the amounts
of upward and downward propagating power must be just that delivered to the
halo across the upper boundary. If that power is all outgoing, the
corresponding result should be: 
\begin{eqnarray}
I_{halo} &=&\frac 12\rho _{hot}v_{hot}^2c_{hot}  \nonumber \\
&\approx &\frac 12P_{hot}v_{hot}^2/c_{hot}  \nonumber \\
&\sim &\frac 12P_{z=0}v_{z=0}^2/c_{hot}  \nonumber \\
&\sim &5\times 10^{-9}\textrm{ erg cm}^{-2}\textrm{ s}^{-1}, \label{eqn7}
\end{eqnarray}
where the next to the last step is taken from the results of our normal mode
analysis and is very approximate. Quantities with subscript \textit{hot} are
evaluated at the lower boundary of the hot region. This result differs from
our upper limit in Eq. \ref{eqn6} by approximately the ratio of the signal 
speed at the midplane to that in the halo. For comparison, the power per unit 
area actually retained by the atmosphere during the simulation (after losses 
due to radiation by the halo, which we estimate to be comparable) is: 
\begin{equation}
I=\frac Et\approx \frac{4\times 10^8\textrm{ erg cm}^{-2}}{3\times 10^9\textrm{
yr}}\approx 4\times 10^{-9}\textrm{ erg cm}^{-2}\textrm{ s}^{-1} .
\label{eqn8}
\end{equation}

Let us compare this with the amount of power released into the Galaxy from
supernovae and spiral arm disturbances. If there is one supernova of energy
10$^{51}$ ergs every 100 years, the average power input into the upper half 
($z\geq 0$) of the Galaxy (assuming a radius of 10 kpc) is roughly: 
\begin{equation}
I\approx \frac 12\frac{10^{51}\textrm{ erg}/100\textrm{ yr}}{\pi \left( 10\textrm{
kpc}\right) ^2}\approx 5\times 10^{-5}\textrm{ erg cm}^{-2}\textrm{ s}^{-1} .
\label{eqn9}
\end{equation}
This is several orders of magnitude greater than the power input of our
perturbations, though it is certainly not provided coherently, or at the
resonant frequency. However spiral wave disturbances, occurring with a period
of perhaps 120 Myr, put at least 3$\times $10$^{-12}$\ ergs per atom into
compression, in a column density of order 3$\times $10$^{20}$\ cm$^{-2}$\
per side. The corresponding average power density per side is roughly: 
\[
I\approx \frac{\left( 3\times 10^{-12}\textrm{ erg}\right) \left( 3\times
10^{20}\textrm{ cm}^{-2}\right) }{1.2\times 10^8\textrm{ yr}}\approx 2.4\times
10^{-7}\ \textrm{erg cm}^{-2}\textrm{ s}^{-1}, 
\]
which is four times the nominal power of our waves in Equation \ref{eqn6}
and sixty times that actually transmitted into our most extreme halo.
Therefore the energy requirement for the resonant production of a hot,
ionized halo above 5 kpc is quite small compared to the energy potentially
available.  The essentially resonant spiral arm period chosen above corresponds to a four 
arm spiral at the solar circle, with a differential speed between gas and 
pattern of 100 km s$^{-1}$.

\section{CONCLUSIONS}

We found several potentially important properties of the interstellar medium
in our experiments. Probably the most intriguing of these is that
segregation of infalling material into two dominant components occurs
naturally after any large expansive event in the ISM as a whole. The
material closest to the midplane returns first and stops. After roughly 50
Myr a substantial low velocity column density has collected, with outer
material continuing to fall in upon it. With a sufficiently large velocity
differential, the two components are separated by a nearly stationary shock.
The large structure known as the IV Arch, covering about half of the
northern sky, could well be the consequence of such a process.

The concentration of material in velocity space (rather than in real space)
is a natural occurrence in the ISM, and can easily lead to apparent
structures in the diffuse gas which mimic low-density clouds. Experiments in
which we added small velocity perturbations (via oscillation of the galactic
midplane) to cases in which a velocity segregated infall was already present
(from a general expansion, as above) provided velocity spectra closely
resembling those of HD 93521, even though there were no density
concentrations, i.e. clouds, whatsoever. We suggest that perhaps only the
one colder component in the spectrum of that star is actually a localized
density feature.

We found the periods of the four lowest normal modes for our model ISM as
functions of the height of the outer boundary, both through Fourier analysis
of the oscillations in the aftermath of a major perturbation, and via linear
analysis, with the two methods giving remarkably similar results. Our
hydrodynamics experiments further revealed two curious tendencies: first
that the oscillating structure appears to concentrate energy into two modes,
which two depending on the nature of the initial perturbation; and second
that the relative amplitudes of the two modes are comparable at the outer
boundary, which therefore shows substantial amplitude modulation in the beat
pattern.

We also explored a timescale that we refer to as the bounce time.
Starting with a large compression of material near the midplane, the bounce
time at a given height is the length of time for the material to return to
its next minimum height. For material in the general vicinity of the plane,
the bounce time tended to lie between the half period of the lowest
frequency breathing mode and the half period of the first overtone,
depending on the magnitude of the initial compression, and therefore perhaps
on the relative amplitudes of the excited modes. In general, the bounce time
for material near the plane lay in the range 40 to 60 Myr, while material
beyond 1.5 kpc took substantially longer to return. It is this
characteristic difference that leads to the velocity segregation discussed
above. It is our impression that the approximate equality between the bounce
time and the half period of the most strongly excited breathing mode derives
from the fact that the amplitude is very nonlinear in the initial and secondary
compressions, with the result that very little time is spent in the
compressed state. Most of the bounce time is spent in the expanded state,
which lasts roughly half the period of the oscillation itself.

For atmospheric boundary heights, $z_{max}$, in the range 2 to 4 kpc, we
found the periods of the fundamental breathing mode to be between 100 and 
120 Myr, and bounce times to be about half those values. Following Martos 
\& Cox (1998), we should then expect to find dense structures within or 
along spiral arms, separated in space by distances proportional to relative
velocity and timescales of 50 to 60 Myr, as the disturbed structure bounces.
With a relative velocity within the arm of 20 km s$^{-1}$, the spacing would
be roughly 1 kpc. With an average relative speed between the pattern and
rotation of 100 km s$^{-1}$, the scale would be 5-6 kpc. If the first
overtone in the breathing mode is excited, the half periods drop to the
range of 30 to 40 Myr, and the characteristic feature spacings for the
assumed velocities reduce to 0.6-0.8 kpc and 3-4 kpc (for relative speeds of 
20 km s$^{-1}$ and 100 km s$^{-1}$, respectively). For comparison, the
circumference at the solar circle is roughly 50 kpc, and the timescale for
material to circumnavigate it, relative to the pattern, is of order 500 Myr.
With a bounce timescale of roughly one tenth that, one might expect to
encounter roughly ten unusually dense regions around the circle, some spaced
closer than 1 kpc (feathers?) with others several kpc
apart along the circle (arms and spurs?). If the full
oscillations of the normal modes (rather than strong bounces) are 
responsible for the structuring,
the feature separations will be roughly twice these, their number on the
solar circle reduced to roughly four or five, possibly reinforcing a four arm
spiral. 

Whether these expectations are borne out in some form by 3-dimensional 
modeling of the arm pattern remains to be seen. Some alteration
is expected as waves which are not entirely vertical will have somewhat
different periods and finite wavelengths parallel to the plane. The basic
idea, however, is that the spiral arm ``shock'' is fragmented, feeding
energy preferentially into over dense regions resulting from waves provoked
by other spiral arm fragments upstream (and slightly off to the side). That
energy input drives further oscillations to provide dense regions further
downstream yet, where the next shock fragments occur. As these shock
fragments are not oriented parallel to the arm locus, significantly stronger
interactions can occur than if their strength were determined only by the
component of the entry velocity normal to that more oblique locus.

An individual large disturbance in the disk leads eventually to a shock wave
propagating up into the low-density outer parts of the atmosphere. This
shock strengthens as it proceeds, finally reaching such low density that it
is no longer radiative on the timescale for propagating through the rest of
the atmosphere. A hot low-density layer is produced, on top of the cool disk
material. With no further disturbance, this layer slowly cools and rejoins
the disk, but the timescale is extraordinarily long. If the time between such 
disturbances is longer than this cooling timescale, then the hot halo may become 
a more or less permanent feature of the atmosphere. Similarly, when low
amplitude waves are generated continuously within the disk and propagate
into its outer reaches, they too can heat an outer layer to high
temperatures, eventually reaching a new equilibrium with the unreflected
portion of the wave power. We found that there is a strong
resonance at which the temperature and size of the outer layer are greatly
enhanced, leading to temperatures in excess of 10$^6$\ K and a thickness of
many kiloparsecs. In our experiments, the period of the resonance was that
of the fundamental breathing mode for the slightly altered disk, essentially
the same oscillation invoked to provide structuring in the spiral arm
pattern. It appears that the two situations are linked. If the spiral waves
are sufficiently large in amplitude to cause substantial vertical motion in
rebounding from their compressions, vertical bounces and subsequent
oscillations will be produced that can both structure the details of the
arms themselves and lead to substantial heating of material at the outer
disk. The latter effect produces a large and very hot gas layer, potentially
contributing to the populations of high stage ions (O~VI in particular) and
to the soft X-ray background. The details of the hot halo depend on the flux
of driving mechanical energy, but quite an impressive halo was formed in our
simulation with a very small power, on the order of 4$\times $10$^{-9}$\ erg 
cm$^{-2}$ s$^{-1}$, which is less that 2\% of that available from spiral
wave disturbances.

Our experiments on hot halo formation were all performed with a nominal disk
boundary height of 5 kpc. The resulting haloes have very low densities, with
thermal pressures more than two orders of magnitude below the total midplane
pressure. We estimate that their O~VI column densities would be of order 10$
^{13}$\ cm$^{-2}$. With a lower boundary height, the halo pressure and O~VI
column densities could be significantly higher, though much more power would
be required to sustain them.

Surely the details of these results will change when the studies are
repeated in two and three dimensions, when MHD effects are included
explicitly rather than just through an effective pressure term, when self
gravity and spiral wave behavior are included, and perhaps again when the
driving terms associated with correlated star formation are incorporated.
But the essential effects are likely to persist. What goes up must come
down. Material nearer the plane returns sooner, hence producing the
``falling sky.'' Velocities associated with small disturbances in the
intercloud medium will produce spectral features, appearing to be clouds
which in reality do not exist. Spiral arm shocks will provoke vertical
motions with periods that will lead to structure within the spiral waves
themselves, and at periods that will tend to promote the formation of a hot
halo overlying the cool disk. It is conceivable that the spiral waves might
also, on occasion, produce such large upflows that the falling sky is an
associated phenomenon. We refer the reader once again to Martos \& Cox
(1998) for further discussion.

\begin{appendix}
\begin{center}
\bf
\appendixname{: Linear Analysis}
\rm
\end{center}

\section{Equation of Motion}

Consider a small parcel initially with position $z_o$, thickness $\Delta z_o$, 
and density $n_o$. After a small disturbance, it has a displacement
$\delta(z_o,t)$,  new position
\begin{equation}
z=z_o+\delta (z_o,t),  \label{eqn10}
\end{equation}
and new thickness 
\begin{equation}
\Delta z=\Delta z_o\left( 1+\frac{\partial \delta (z_o,t)}{\partial z_o}
\right) =\Delta z_o\left( 1+\delta ^{\prime }\right) , \label{eqn11}
\end{equation}
where we have introduced the notation $\delta ^{\prime }=\partial \delta
(z_o,t)/\partial z_o$. From continuity, we have $n\Delta z=n_o\Delta z_o$,
so: 
\begin{equation}
\frac \rho {\rho _o}=\frac n{n_o}=\frac{\Delta z_o}{\Delta z}=\frac
1{1+\delta ^{\prime }}\approx 1-\delta ^{\prime } .  \label{eqn12}
\end{equation}

The (comoving) velocity of the parcel is: 
\begin{equation}
v=\frac{Dz}{Dt}=\frac D{Dt}\left( z_o+\delta \right) =\frac{\partial \delta 
}{\partial t}=\dot{\delta} , \label{eqn13}
\end{equation}
with $\dot{\delta}=\partial \delta (z_o,t)/\partial t$. Its gravitational
acceleration at its perturbed location is: 
\begin{equation}
g=g_o+\frac{\partial g}{\partial z_o}\delta =g_o+g^{\prime }\delta ,
\label{eqn14}
\end{equation}
where $g_o=g(z_o)$.

Assuming that the gas remains nearly isothermal (as in our
hydrodynamic simulations), the new pressure of the parcel is: 
\begin{eqnarray}
P &=&\chi nkT+\frac{B^2}{8\pi }  \nonumber \\
&=&\chi n_okT\left( \frac n{n_o}\right) +\frac{B_o^2}{8\pi }\left( \frac
n{n_o}\right) ^2  \nonumber \\
&=&P_{To}\left( \frac n{n_o}\right) +P_{Bo}\left( \frac n{n_o}\right) ^2 
\nonumber \\
&\approx &\left( P_{To}+P_{Bo}\right) -\left( P_{To}+2P_{Bo}\right) \delta
^{\prime }  \nonumber \\
&\approx &P_o-\left( \rho _oc_o^2\right) \delta ^{\prime }, \label{eqn15}
\end{eqnarray}
where $P_{To}=\chi n_okT$ is the initial thermal pressure, 
$P_{Bo}=B_o^2/8\pi$ is the initial magnetic pressure, $P_o=P_{To}+P_{Bo}$ is
the total initial pressure, $(n/n_o)^q\approx 1-q\delta ^{\prime }$, and the 
signal speed (i.e. magnetosonic velocity) is given by: 
\begin{equation}
c^2=\frac{\partial P}{\partial \rho } .  \label{eqn23}
\end{equation}
So the equilibrium signal speed, $c_o$, is given by: 
\begin{equation}
c_o^2=\frac{P_{To}+2P_{Bo}}{\rho _o} .
\end{equation}

The pressure gradient is then: 
\begin{eqnarray}
\frac{\partial P}{\partial z}&=&\frac{\partial z_o}{\partial z}\frac{
\partial P}{\partial z_o}  \nonumber \\
&=& \left(\frac{1}{1+\delta ^{\prime}}\right) \left(P_o - \left(\rho_o
c_o^2\right) \delta ^{\prime}\right)^{\prime}  \nonumber \\
&=& \left(\frac{\rho}{\rho_o}\right) \left(P_o^{\prime} - \left(\rho_o
c_o^2\right)^{\prime} \delta ^{\prime}- \left(\rho_o c_o^2\right) \delta
^{\prime\prime}\right) .  \label{eqn16}
\end{eqnarray}

Substituting these results into Euler's equation and then canceling the
equilibrium terms, we get: 
\begin{eqnarray}
\frac{Dv}{Dt} &=&-\frac 1\rho \frac{\partial P}{\partial z}-g \nonumber \\
\frac D{Dt}\dot{\delta} &=&-\frac 1\rho _o\frac{\partial P_o}{\partial z_o}+
\frac{\delta ^{\prime }}{\rho _o}\frac d{dz_o}\left( \rho _oc_o^2\right)
+\delta ^{\prime \prime }c_o^2-\left( g_o+g^{\prime }\delta \right)\nonumber \\
\ddot{\delta} &=&-g^{\prime }\delta +\frac 1{\rho _o}\left( \rho
_oc_o^2\right) ^{\prime }\delta ^{\prime }+c_o^2\delta ^{\prime \prime } .
\label{eqn26}
\end{eqnarray}
Equation \ref{eqn26} is the general equation of motion for linear waves and
oscillations. The input information required includes $g(z)$, $\rho _o(z)$, $
T$, $\chi $, and the location of and pressure at the outer boundary. From those 
$g(z)^{\prime }$, $P_T(z)$, $P_B(z)$, and then $c_o^2(z)$\ can be found as
per the prescription in the text. In the next two sections, we consider
first high frequency waves to learn their amplitude dependence on $z$, and
then normal modes.

\section{High Frequency Waves}

Assuming a harmonic disturbance of the form: 
\begin{equation}
\delta \left( z,t\right) =A\left( z\right) \exp \left[ i\left( kz-\omega
t\right) \right] , \label{eqn27}
\end{equation}
we can force most of the $z$ dependence into the wave number $k$, and solve
for the gradual change in the amplitude $A\left(z\right)$. 
Equation \ref{eqn26} becomes: 
\begin{equation}
-\omega ^2A=-g^{\prime }A+\frac 1{\rho _o}\left( \rho _oc_o^2\right)
^{\prime }\left( A^{\prime }+ikA\right) +c_o^2\left[ \left( A^{\prime \prime
}-k^2A\right) +i\left( 2kA^{\prime }+k^{\prime }A\right) \right] .
\label{eqn29}
\end{equation}

In order to satisfy this equation for all time and space, the real and
imaginary parts must each be satisfied. From the imaginary part, we have: 
\begin{eqnarray}
\frac{\left( \rho _oc_o^2\right) ^{\prime }}{\rho_oc_o^2}+2\frac{A^{\prime } 
}A+\frac{k^{\prime }}k &=&0  \nonumber \\
\rho _oc_o^2A^2k &=&const.  \label{eqn30}
\end{eqnarray}
The real part yields: 
\begin{eqnarray}
-\omega ^2A &=&-g^{\prime }A+\frac 1{\rho _o}\left( \rho _oc_o^2\right)
^{\prime }A^{\prime }+c_o^2\left( A^{\prime \prime }-k^2A\right) \nonumber \\
-\omega ^2 &=&-g^{\prime }+\frac 1{\rho _o}\left( \rho _oc_o^2\right)
^{\prime }\frac{A^{\prime }}A+c_o^2\frac{A^{\prime \prime }}A-c_o^2k^2 .
\label{eqn31}
\end{eqnarray}
Using the results from the imaginary part, we can write this as: 
\begin{equation}
-\omega ^2=-g^{\prime }+\frac 1{\rho _o}\left( \rho _oc_o^2\right) ^{\prime
}\left\{ -\frac 12\left[ \frac{k^{\prime }}k+\frac{\left( \rho
_oc_o^2\right) ^{\prime }}{\rho _oc_o^2}\right] \right\} +c_o^2\frac{
A^{\prime \prime }}A-c_o^2k^2 .  \label{eqn32}
\end{equation}
By high frequency waves, we mean those whose wavelength ($\lambda $) is
sufficiently small that 
\begin{equation}
\frac{\lambda ^2}{h^2}\ll 1\textrm{, and thus} \frac{A^{\prime \prime }}A \sim
\frac1{h^2}\ll k^2 , \label{eqn33}
\end{equation}
where $h$ is the functional scale height of the atmosphere. With this
assumption the middle two terms of Equation \ref{eqn32}, which are second
order, can be neglected, giving: 
\begin{equation}
\omega ^2=g^{\prime }+c_o^2k^2 .  \label{eqn34}
\end{equation}
For high frequencies, we can also neglect the $g^{\prime }$ $(=-\phi
^{\prime \prime })$ term, giving us: 
\begin{equation}
\omega \approx c_ok . \label{eqn35}
\end{equation}
Thus Equation \ref{eqn30} becomes: 
\begin{equation}
\rho _oc_o^2A^2k=\rho_oc_oA^2\omega \cong const.  \label{eqn36}
\end{equation}
Since the frequency is constant, we have finally: 
\begin{equation}
\rho _oc_oA^2\approx const.  \label{eqn37}
\end{equation}

Equation \ref{eqn37} is the standard result introduced in Section \ref
{secSmallOsc} for the growth in amplitude of short wavelength waves. The
amplitude is proportional to $(\rho_o c_o)^{-1/2}$, where $\rho_o$\ and
$c_o$\ are the local unperturbed density and signal speed as functions of $z$.

\section{Normal Modes}

Equation \ref{eqn26} can be simplified for use in searching for the normal
modes of the disk by assuming that $\delta $ is a product of two functions: 
\begin{equation}
\delta \left( z_o,t\right) =f\left( z_o\right) \epsilon \left( z_o,t\right).
\label{eqn38}
\end{equation}
Then equation \ref{eqn26} becomes: 
\begin{equation}
\frac{\ddot{\epsilon}}{c_o^2} =\frac{\left( \rho _oc_o^2\right) ^{\prime }
}{\rho _oc_o^2}\left[ \left( \frac{f^{\prime }}f\right) \epsilon +\epsilon
^{\prime }\right] +\left[ \left( \frac{f^{\prime }}f\right) ^2+\left( \frac{
f^{\prime }}f\right) ^{\prime }\right] \epsilon +2\left( \frac{f^{\prime }}
f\right) \epsilon ^{\prime }+\epsilon ^{\prime \prime }-\frac{g^{\prime }}{
c_o^2}\epsilon.
\label{eqn42}
\end{equation}

We can make the $\epsilon ^{\prime }$ terms cancel if we define: 
\begin{equation}
F\equiv \left( \frac{f^{\prime }}f\right) =-\frac 12\frac{\left( \rho
_oc_o^2\right) ^{\prime }}{\rho _oc_o^2} .  \label{eqn43}
\end{equation}
so equation \ref{eqn42} becomes: 
\begin{equation}
\frac{\ddot{\epsilon}}{c_o^2}  = \epsilon ^{\prime \prime }-\left( \frac{
g^{\prime }}{c_o^2}+F^2-F^{\prime }\right) \epsilon . \label{eqn44}
\end{equation}
Further defining $K_o^2$ such that: 
\begin{equation}
K_o^2\left( z_o\right) \equiv \frac{g^{\prime }\left( z_o\right) }{
c_o^2\left( z_o\right) }+F^2\left( z_o\right) -F^{\prime }\left( z_o\right),
 \label{eqn45}
\end{equation}
and invoking
\begin{equation}
\ddot{\epsilon}=-\omega ^2\epsilon , \label{eqn47}
\end{equation}
in our search for normal modes, we have
\begin{equation}
\epsilon ^{\prime \prime }=-\left[ \frac{\omega ^2}{c_o^2\left( z_o\right) }
-K_o^2\left( z_o\right) \right] \epsilon  . \label{eqn48}
\end{equation}
This is a second order differential equation with no first derivative term,
similar to the time independent Schr\"odinger Equation. It can easily be
solved numerically, and the eigenvalues $\omega $\ searched for such that
the boundary conditions are satisfied.  The outer boundary condition,
however, is most easily described in terms of the original variable $\delta$, 
to which we return next.  (For an isothermal atmosphere with 
constant gravity, the quantity in brackets above is constant and the solutions 
sinusoids, providing a nice analytic test of numerical methods.  Similarly, 
an infinite isothermal atmosphere with linear gravity has polynomial solutions 
like the quantum mechanical harmonic oscillator, the eigenvalues of which 
provide the mode frequencies in the limit of large-scale height, another useful test.)  

Equation \ref{eqn43} implies: 
\begin{equation}
f^2\rho _oc_o^2=const.  \label{eqn53}
\end{equation}
At the midplane, where $\rho _oc_o^2=\rho _{oo}c_{oo}^2$, we define $f(0)=1$.
Hence: 
\begin{equation}
f=\sqrt{\frac{\rho _{oo}c_{oo}^2}{\rho _o\left( z_o\right) c_o^2\left(
z_o\right) }} ,  \label{eqn54}
\end{equation}
and the displacement is given by: 
\begin{equation}
\delta =\epsilon \left( z_o\right) \sqrt{\frac{\rho _{oo}c_{oo}^2}{\rho
_o\left( z_o\right) c_o^2\left( z_o\right) }} .  \label{eqn55}
\end{equation}

Identification of the appropriate outer boundary condition is difficult,
but we have designed the last parcel of the hydrocode to have only thermal
pressure within it, and to have effectively zero pressure on its outer
boundary. We therefore suspected that the appropriate boundary condition for
the linear analysis might be that of a constant zero pressure. This would
then require having zero change in the pressure, a pressure node, at the
outer edge. The comoving pressure change is due to a change in density,
such that $\Delta P=c_o^2\Delta \rho $, or with Equation \ref{eqn15}: 
\begin{eqnarray}
\Delta P &=&-\rho _oc_o^2\delta ^{\prime }=0\textrm{ at }z=z_{\max }  \nonumber
\\
&\Rightarrow &\delta ^{\prime }\left( z_{\max }\right) =0  . \label{eqn56}
\end{eqnarray}
Hence the outer boundary would be an antinode in $\delta $, with no
variation in pressure or density, and an extremum in the velocity. Referring
to Figures \ref{fig4} and \ref{fig8}, it is easy to see why we were
discouraged with this assumption. The pressures and densities do vary
considerably in the outer parcels of the code. Sometimes, but not
frequently, the velocity has an extremum there. And yet, continuing with the
assumption that $\delta $\ should have an antinode at the boundary
gave the same set of frequency eigenvalues found
hydrodynamically, while no other assumption came close.

The boundary condition at the midplane depends on whether the velocity is an
even or odd function of $z$. In terms of our other functions, the velocity
is: 
\begin{equation}
v\left( z_o\right)  = \dot{\delta}\left( z_o\right) =f\left( z_o\right) \dot{
\epsilon}\left( z_o\right) 
\label{eqn57}
\end{equation}
Because $f$ is an even function of $z$, even solutions in $v$\ require: 
\begin{equation}
\epsilon ^{\prime }\left( 0\right) =0  . \label{eqn60}
\end{equation}
These are the ``sloshing'' modes, symmetric in $v$\ and $\delta $\ and
antisymmetric in $\rho$.

For odd solutions in $v$: 
\begin{equation}
\epsilon \left( 0\right) =0  . \label{eqn62}
\end{equation}
These are the ``breathing'' modes, antisymmetric in $v$\ and $\delta$\ and
symmetric in $\rho$.

Our procedure for solving for the normal modes involved, for a chosen value
of $\omega$, integrating Equation \ref{eqn48} upward from $z = 0$ where the
boundary condition was taken to be either $\epsilon \left( 0\right) =1$\ and $
\epsilon ^{\prime }\left( 0\right) =0$\ for the sloshing modes, or $
\epsilon \left( 0\right) =0$\ and $\epsilon ^{\prime }\left( 0\right) =1$\
for the breathing modes, with 1 being chosen as an arbitrary nonzero value.
We also evaluated Equation \ref{eqn55} for $\delta \left(z\right)$, and then
scanned in $\omega$\ until the outer boundary condition of Equation \ref
{eqn56} was satisfied with either zero or one node between the midplane and
the outer boundary, finding in this way the fundamental and first overtones
for that symmetry. As reported in the text, our results for the allowed
periods were nearly identical to those found from the hydrodynamics.

\end{appendix}

\acknowledgements

This work was supported in part by NASA grants NAG5-3155 and NAG5-8417 to
the University of Wisconsin-Madison.


\clearpage

\figcaption[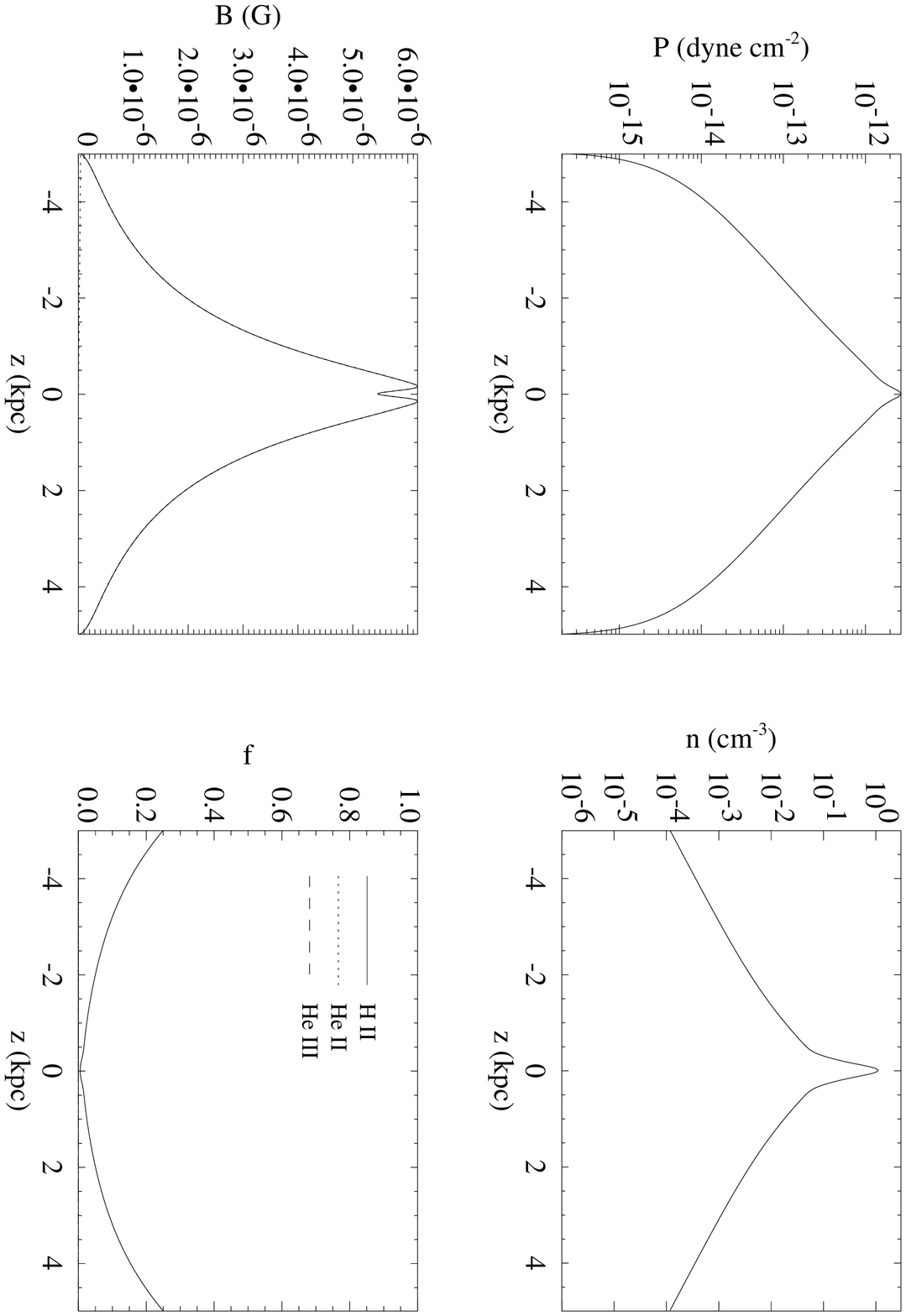]{Initial (equilibrium) distributions of
pressure, number density, magnetic field, and ion fractions for a two-sided,
5 kpc high disk. The slight dip in magnetic field at the midplane is an 
artifact of assuming a uniform temperature of 10$^4$ K, which is slightly too 
high for the smoothed cloud components near the midplane. (Note that the lines 
for He II and He III are not visible because helium is essentially neutral at 
the start of the simulation.)
\label{fig1}}

\figcaption[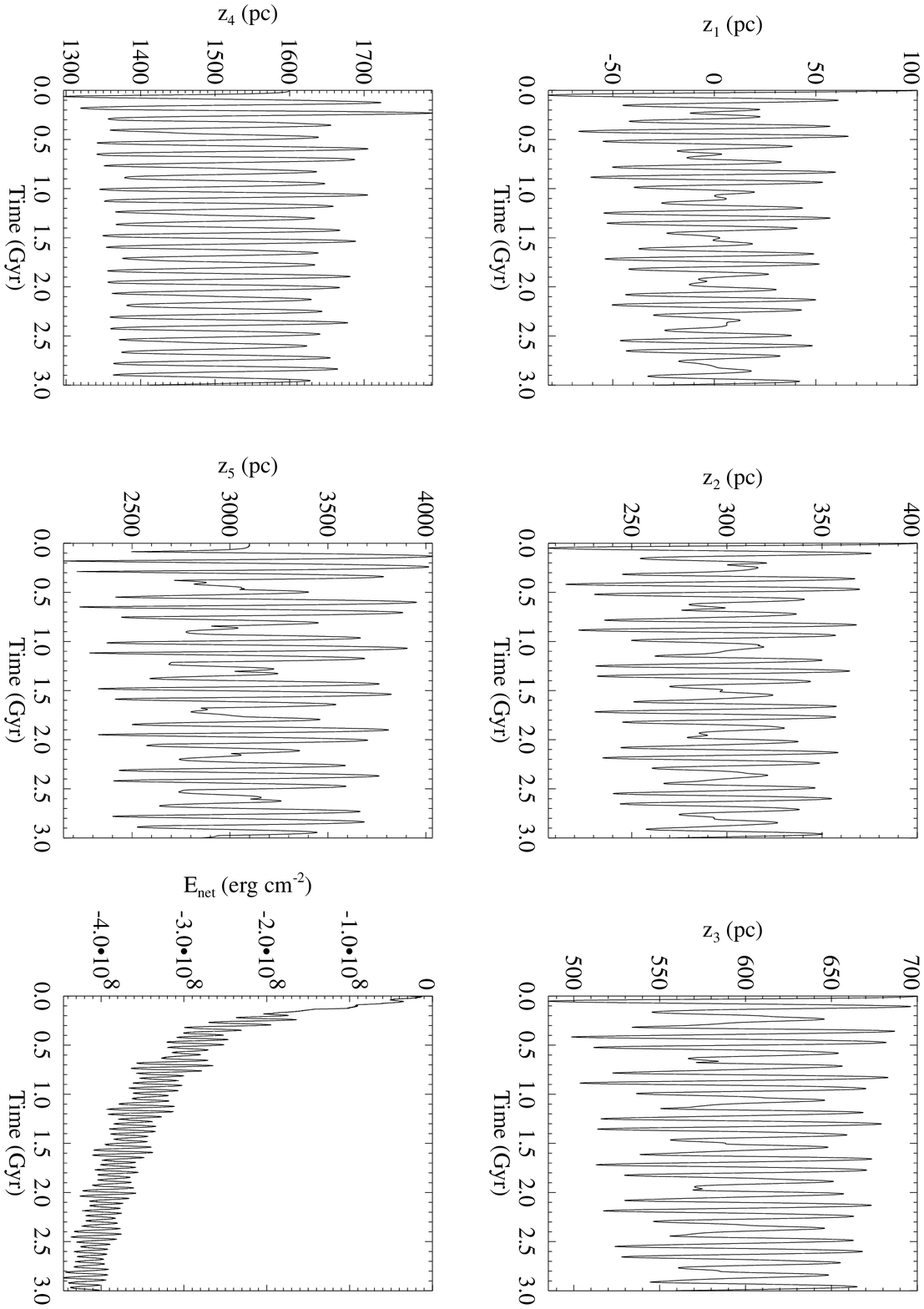]{Positions of tracer parcels and net energy vs. time
following a uniform +100 pc displacement. These plots show, as a function of
time, the locations of the (upper boundaries of) 5 parcels chosen to
illustrate the motions of the disk. Their equilibrium heights are (starting
from the upper left) 0, 300 pc, 600 pc, 1500 pc, and 3000 pc. Also shown, in
the lower right plot, is the change in the total energy (per unit area) in 
the model over time. Note that all of these plots show signs of periodic 
behavior, which is due to excited sloshing modes.
\label{fig2}}

\figcaption[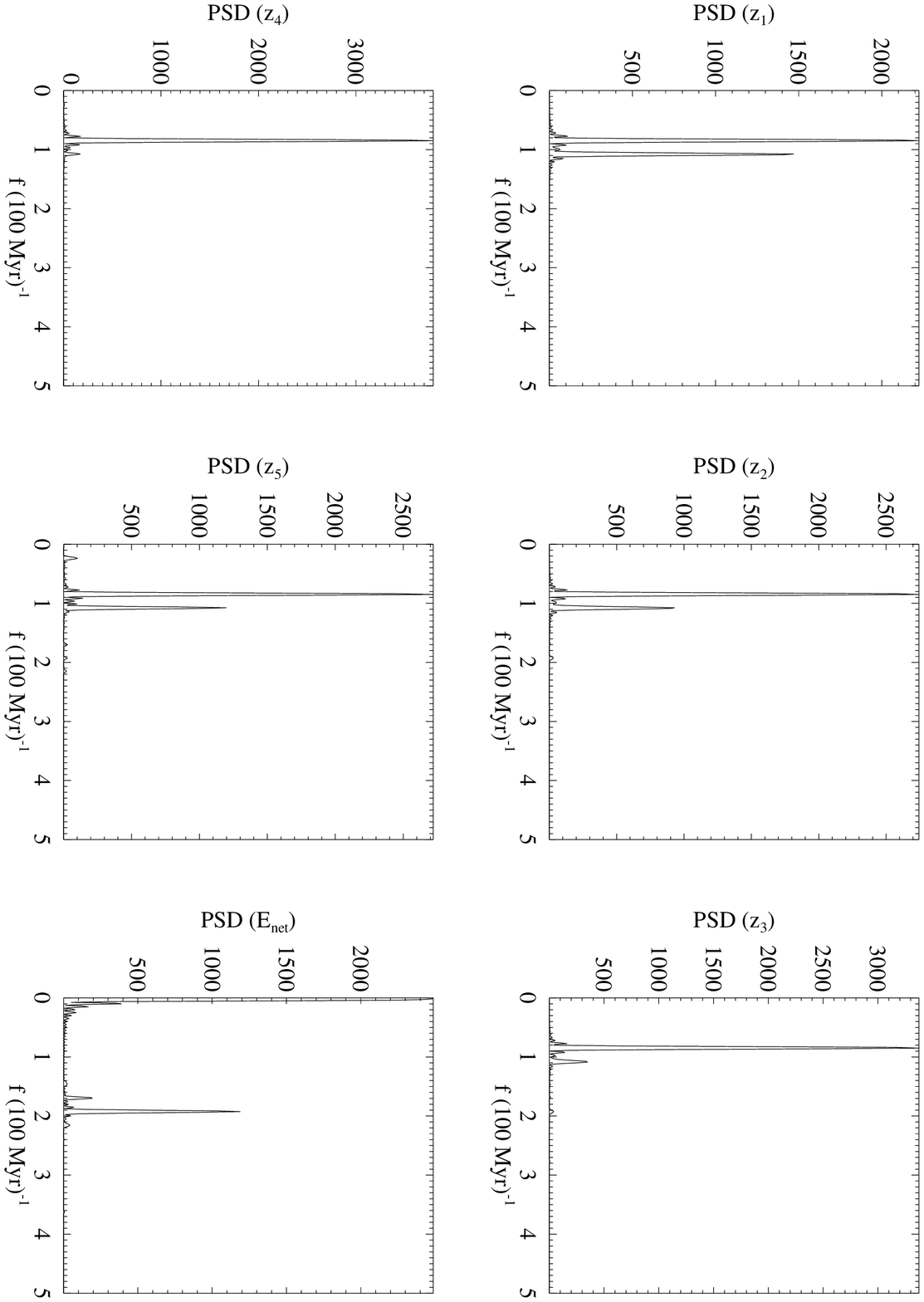]{Fourier transforms of the histories shown in
Figure \ref{fig2}. Each plot is the power spectrum of the Fourier transform
of the corresponding record shown in Figure \ref{fig2}. To eliminate
transients, only the last 2 Gyr of each record were used. Note that the
peak frequencies are the same for all the parcels. The transform of the energy gives
twice the fundamental frequency because it includes \textit{both} sides
of the atmosphere.
\label{fig3}}

\figcaption[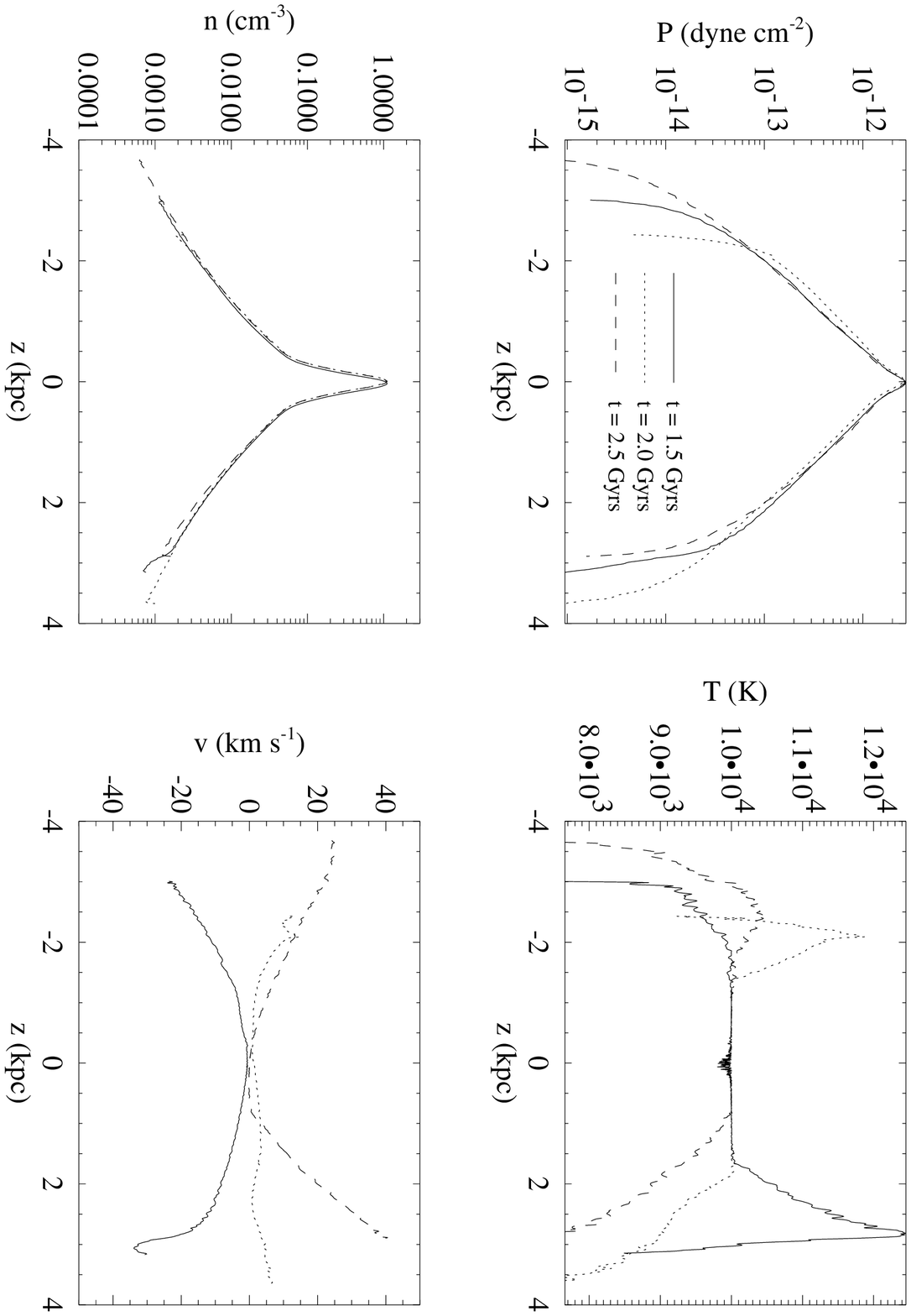]{Distributions of pressure, temperature, density,
and velocity at 1.5, 2.0, and 2.5 Gyr after the equilibrium atmosphere has
been subjected to a large asymmetric disturbance. In the linear
approximation, the velocity profile would be symmetric for pure sloshing modes.
Deviations from that symmetry are apparent and are discussed in the text.
The pressure profile at 2 Gyr shows displacement in opposite directions at
high vs. low $z$, due to the presence of the first overtone.
\label{fig4}}

\figcaption[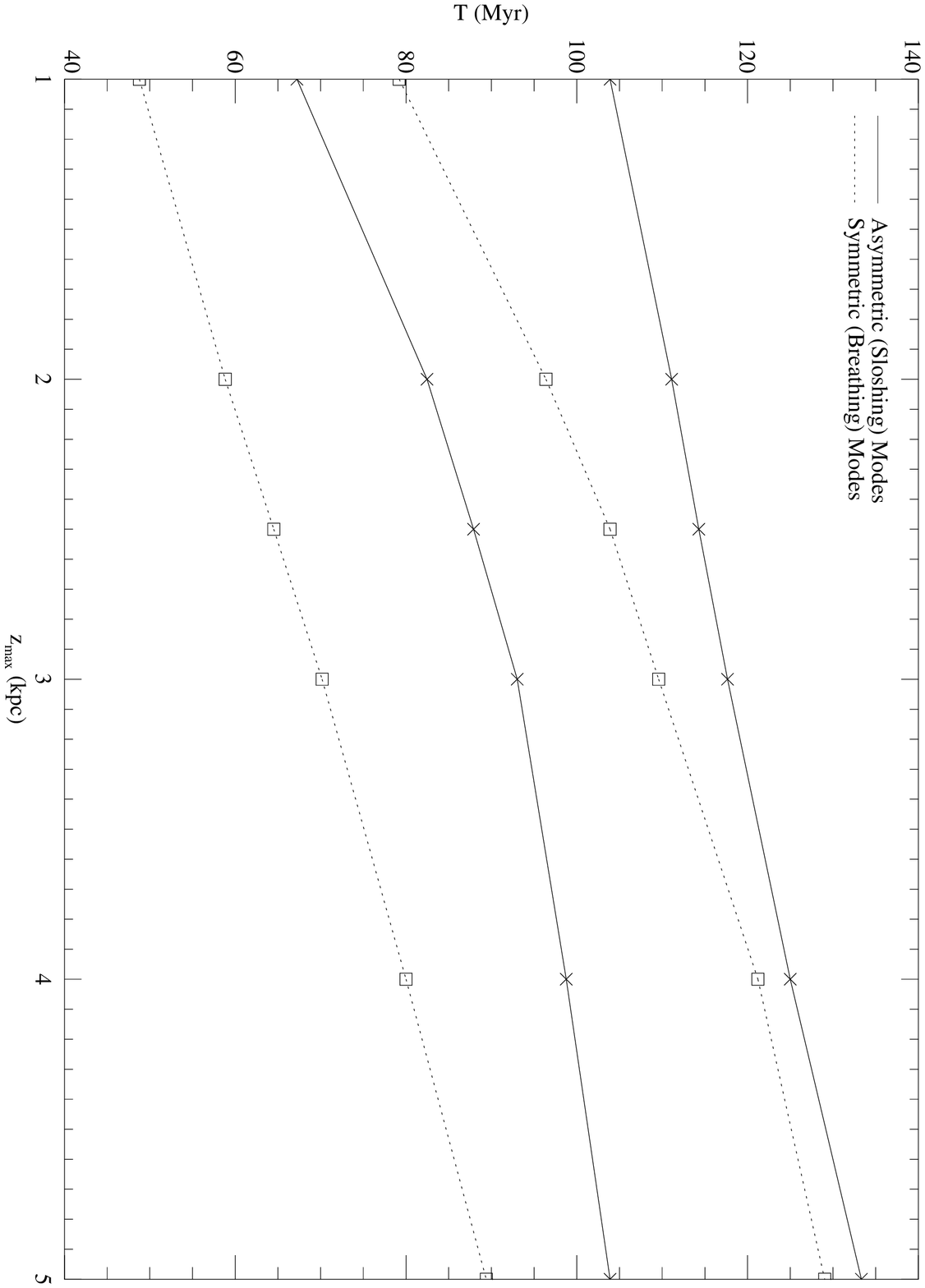]{Dependence of the periods of the asymmetric
(sloshing) and symmetric (breathing) modes on the boundary height of the
model atmosphere.
\label{fig5}}

\figcaption[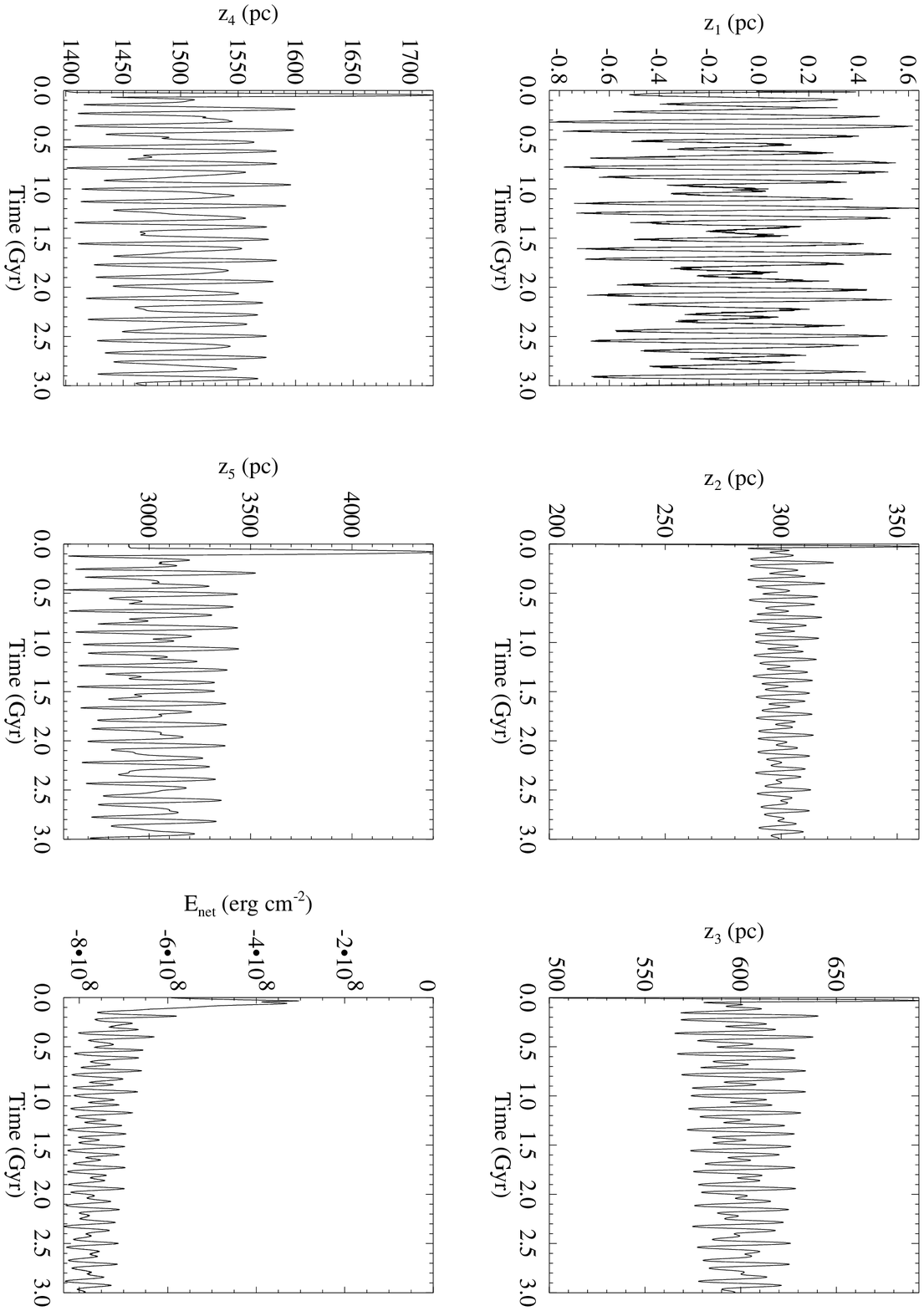]{Heights of five tracer parcels and, in the lower
right, net energy change vs. time following a compression (by a factor of
1.5) of the atmosphere near the midplane. As before, the equilibrium heights 
of the parcels are 0, 300 pc, 600 pc, 1500 pc, and 3000 pc. Again, all of the 
plots show signs of periodic behavior, which in this case is the result of 
excited breathing modes.
\label{fig6}}

\figcaption[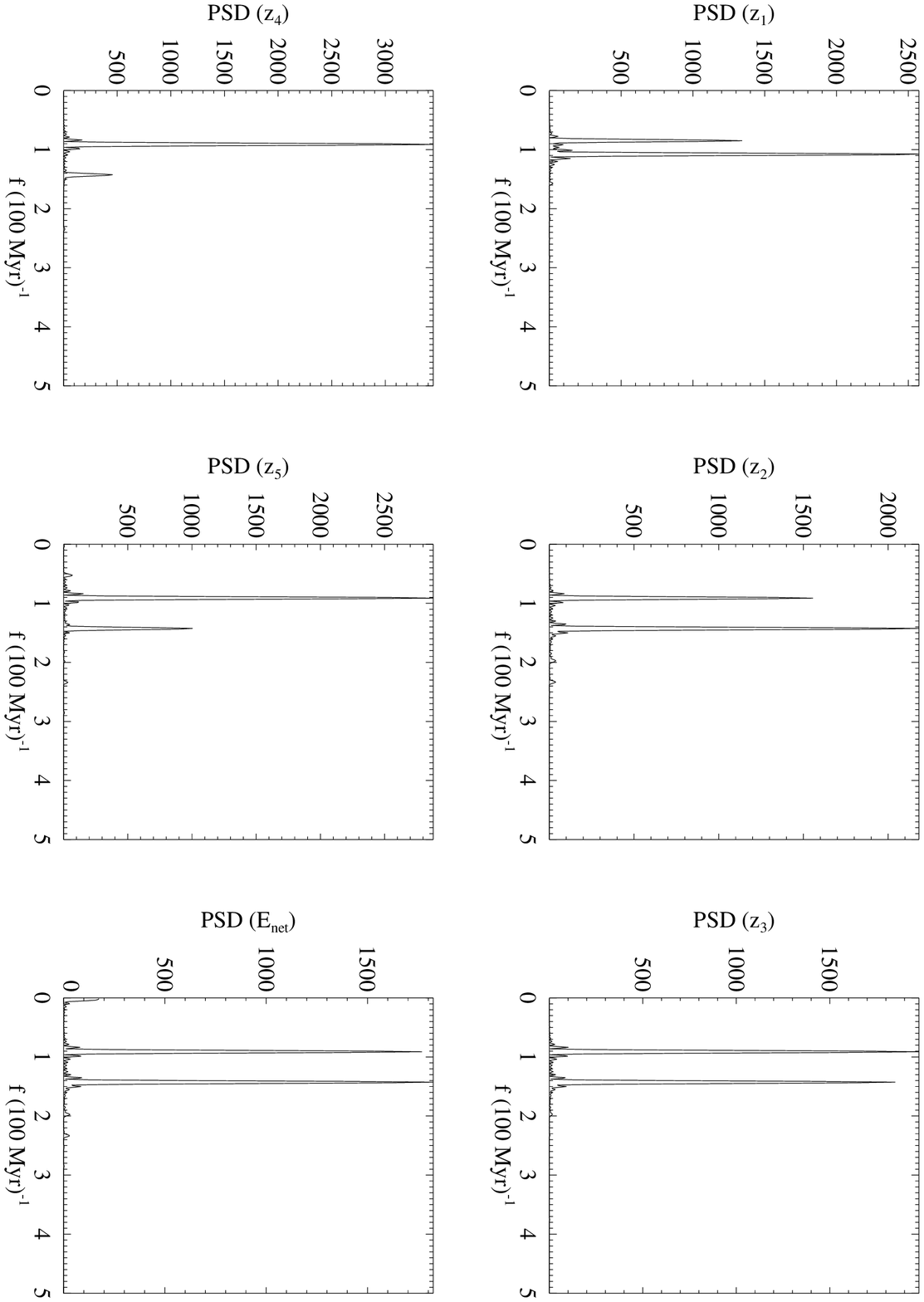]{Fourier transforms of the histories shown in
Figure \ref{fig6}. Each plot is the power spectrum of the Fourier transform
of the corresponding record shown in Figure \ref{fig6}. To eliminate transients,
only the last 2 Gyr of each record was used. The periods are the same for
all the parcels, except at the midplane, where the periods are those of the 
asymmetric (sloshing) modes. These modes, which are probably excited by round-off
errors, are the only ones possible at the midplane because there is no way for the 
midplane to oscillate symmetrically.
\label{fig7}}

\clearpage

\figcaption[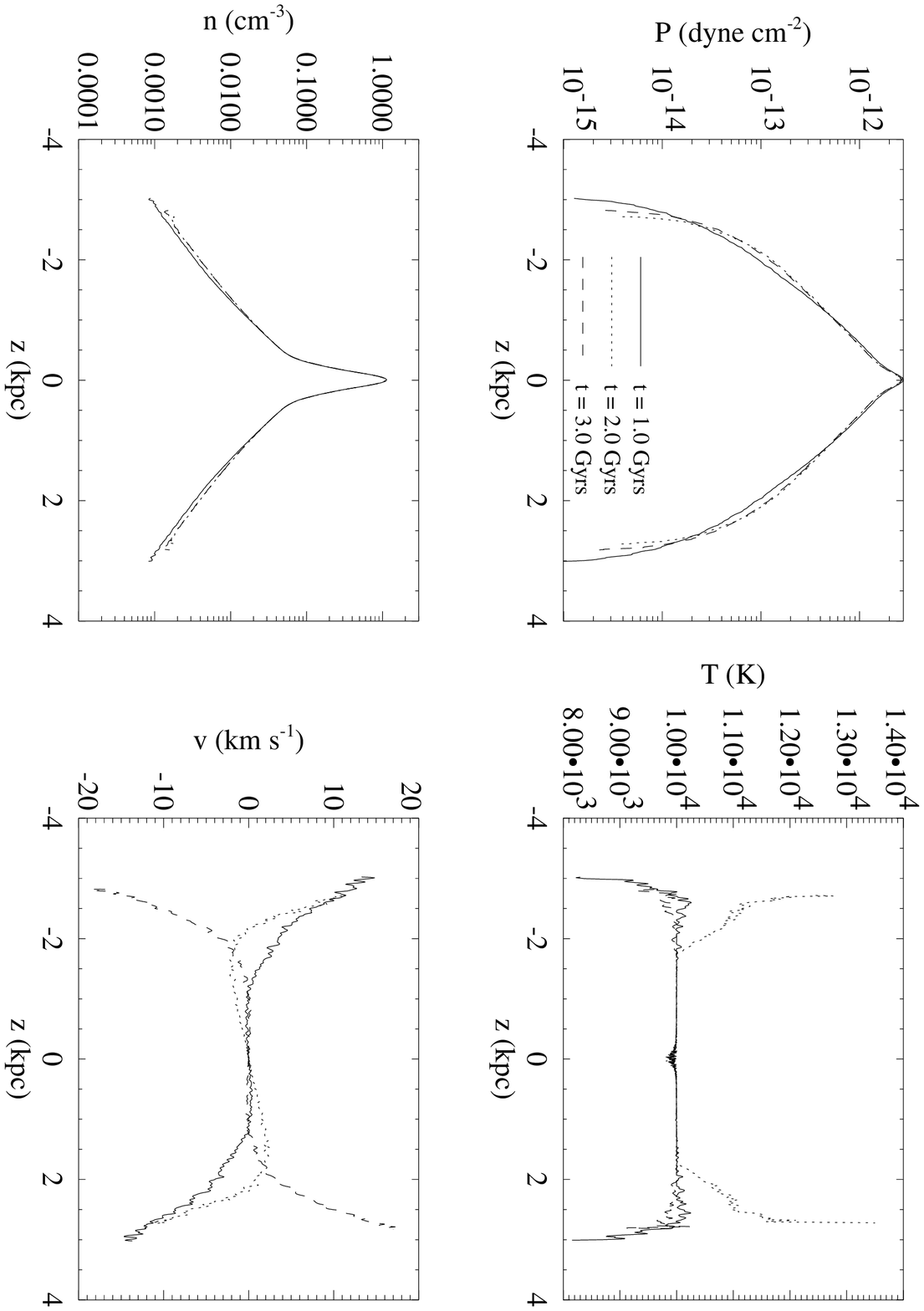]{Distributions of pressure, temperature, density,
and velocity at 1, 2, and 3 Gyr after the equilibrium atmosphere has been
subjected to a large symmetric disturbance. The pressure, density, and
temperature profiles are very symmetric about the midplane, while the
velocities are antisymmetric as in the linear approximation, despite the
small admixture of the fundamental sloshing mode revealed by the motion at
the midplane.
\label{fig8}}

\figcaption[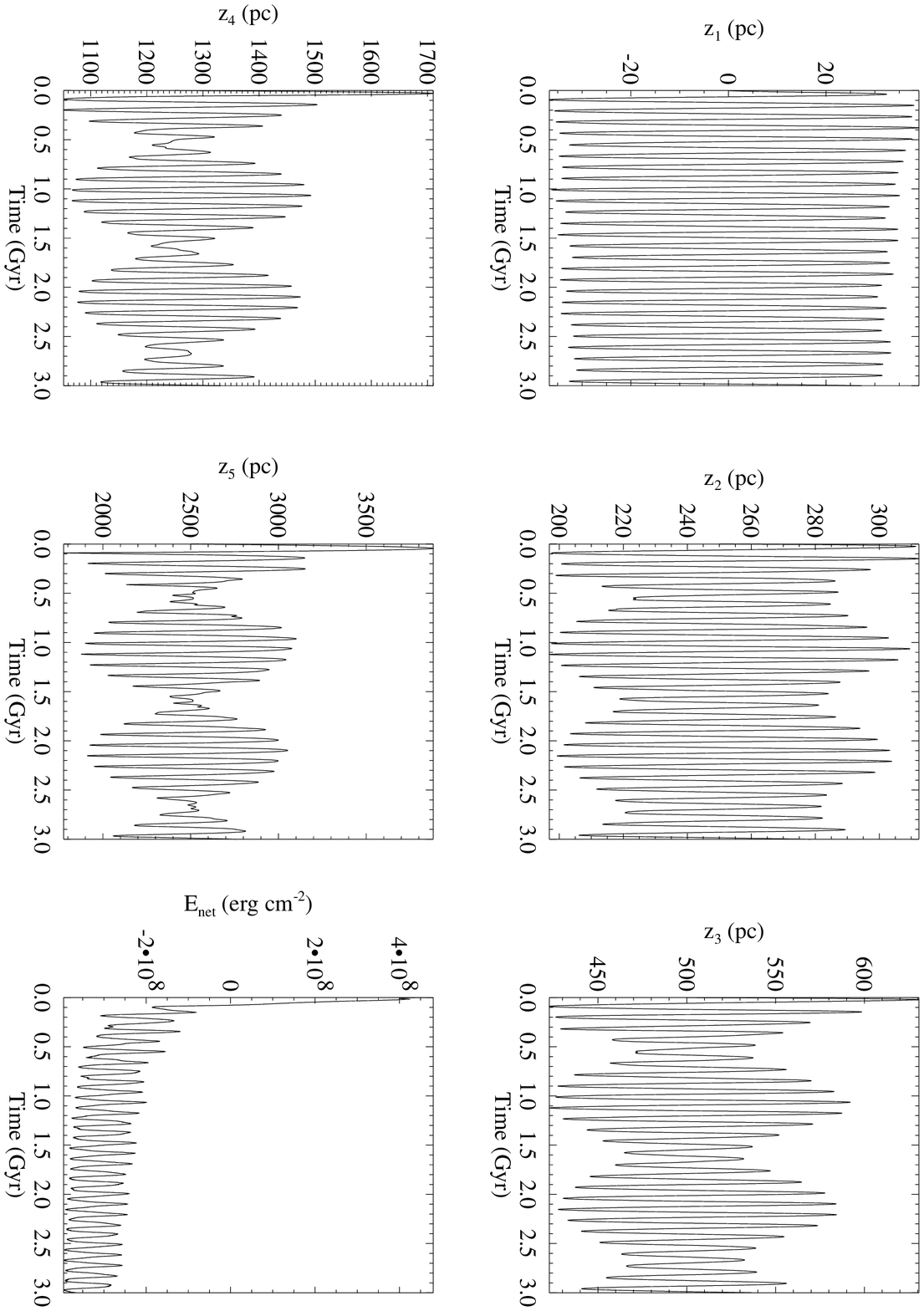]{Heights of tracer parcels and net energy vs. time
for a disturbance that excites both asymmetric and symmetric oscillations.
In this case, the height of the atmosphere is 2.5 kpc, and the tracer
parcels have equilibrium positions of 0, 250 pc, 500 pc, 1250 pc, and 2500
pc. The beating between the two modes is clearly evident, increasing to full
modulation at the top of the atmosphere.
\label{fig9}}

\figcaption[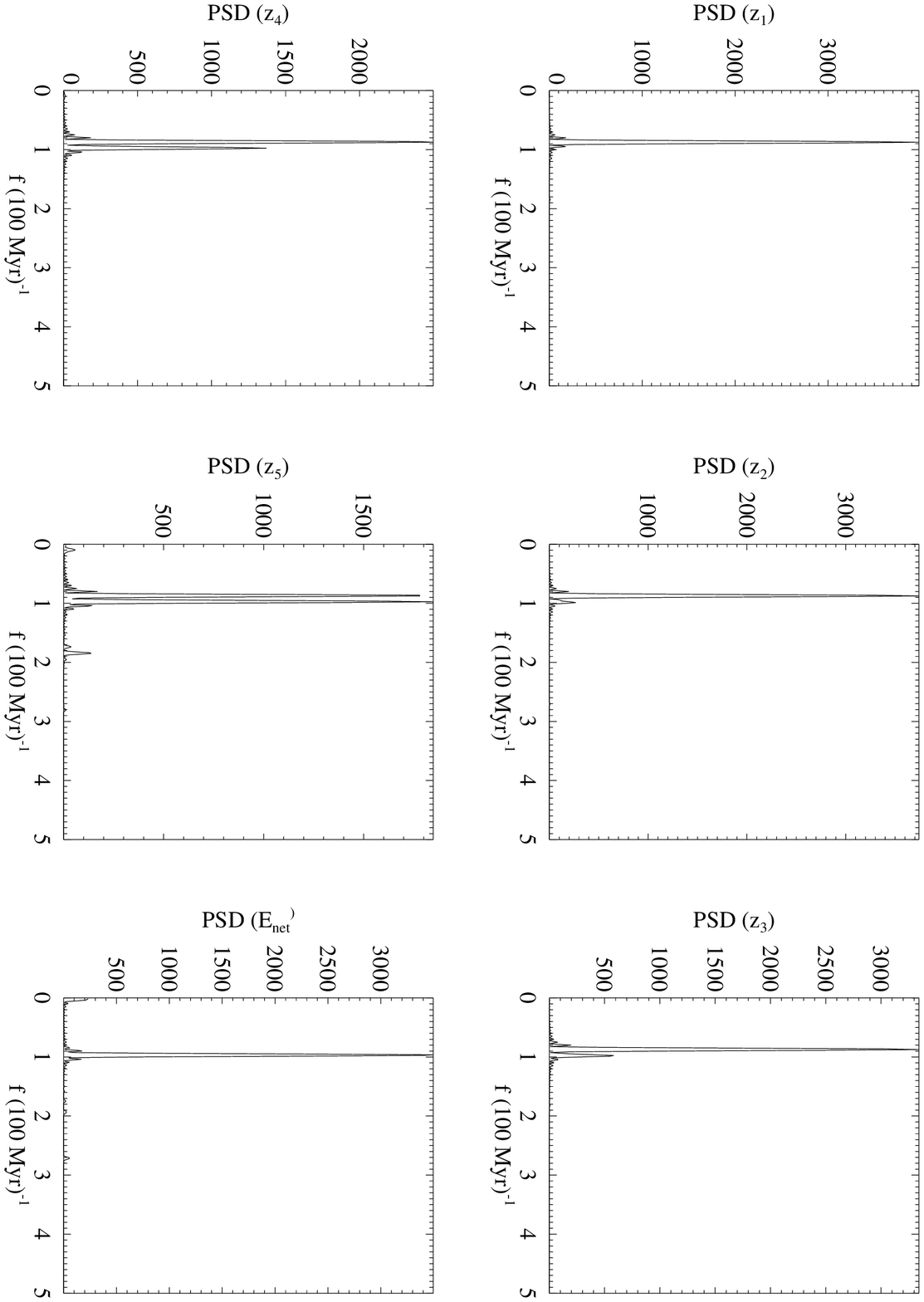]{Fourier transforms of the histories in Figure 
\ref{fig9}. Each plot is the power spectrum of the Fourier transform of the
corresponding record shown in Figure \ref{fig9}. To eliminate transients,
only the last 2 Gyr of each record was used. Note that all of the parcels
except the one at the midplane show two closely spaced frequencies
corresponding to the first asymmetric (sloshing) and symmetric (breathing) 
modes, the relative
amplitude of the latter increasing with $z$. The midplane shows only the
asymmetric mode, and the energy shows only the symmetric mode.
\label{fig10}}

\figcaption[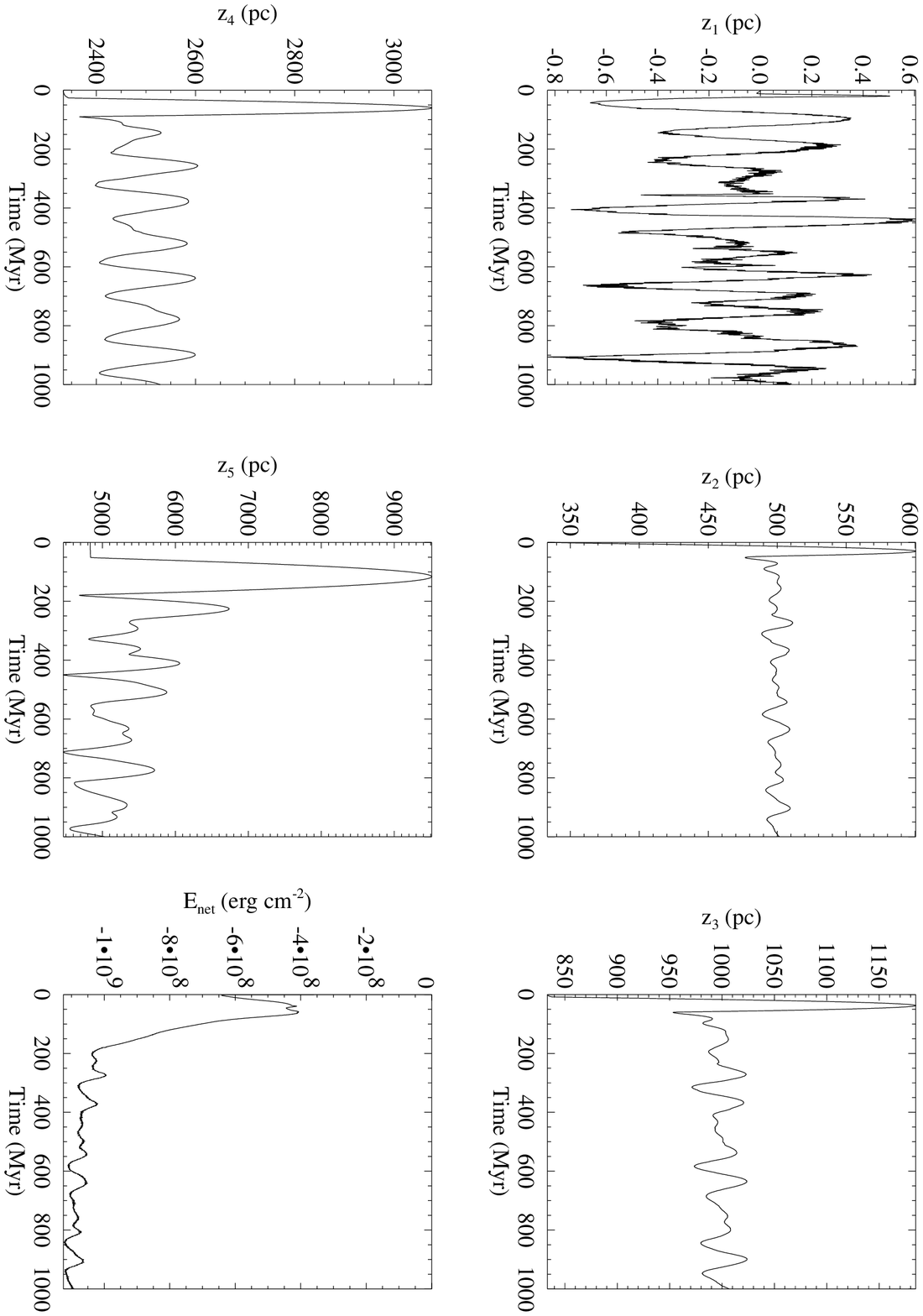]{Positions of tracer parcels and net energy vs. time
following a compression (by a factor of 1.5) of the lower atmosphere. The
height of the atmosphere is 5 kpc and the equilibrium heights of the tracer
parcels are 0, 500 pc, 1000 pc, 2500 pc, and 5000 pc. Only the first 1 Gyr
of the simulation is shown. Note that the transient response is much larger
than the subsequent oscillations.
\label{fig11}}

\figcaption[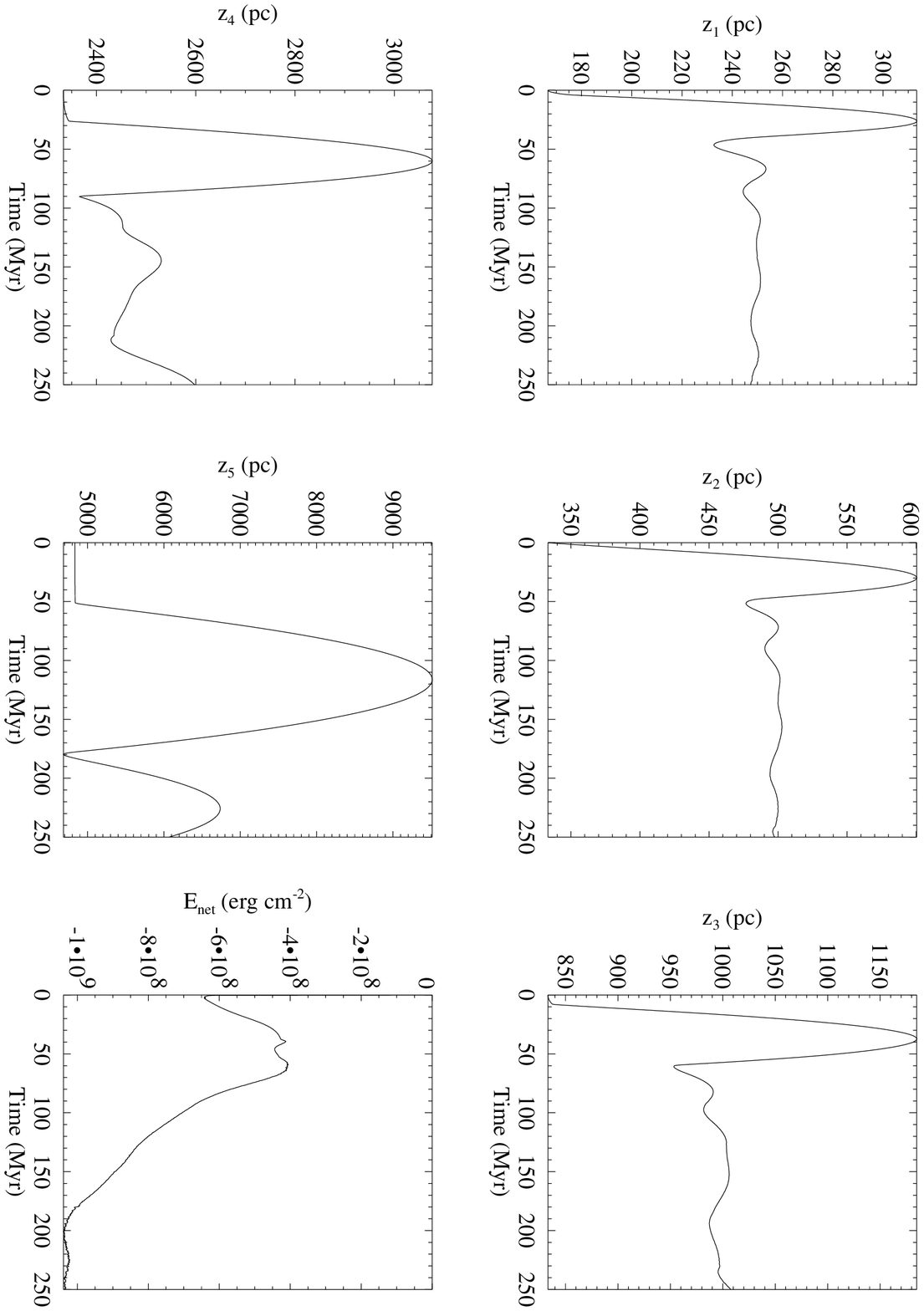]{Same as Figure \ref{fig11}, except with the first
tracer parcel at an equilibrium height of 250 pc and showing only the first
250 Myr of the simulation. Note that the timescale of the transient response
is about the same for the first three tracer parcels, which are in the lower
half of the atmosphere.
\label{fig12}}

\figcaption[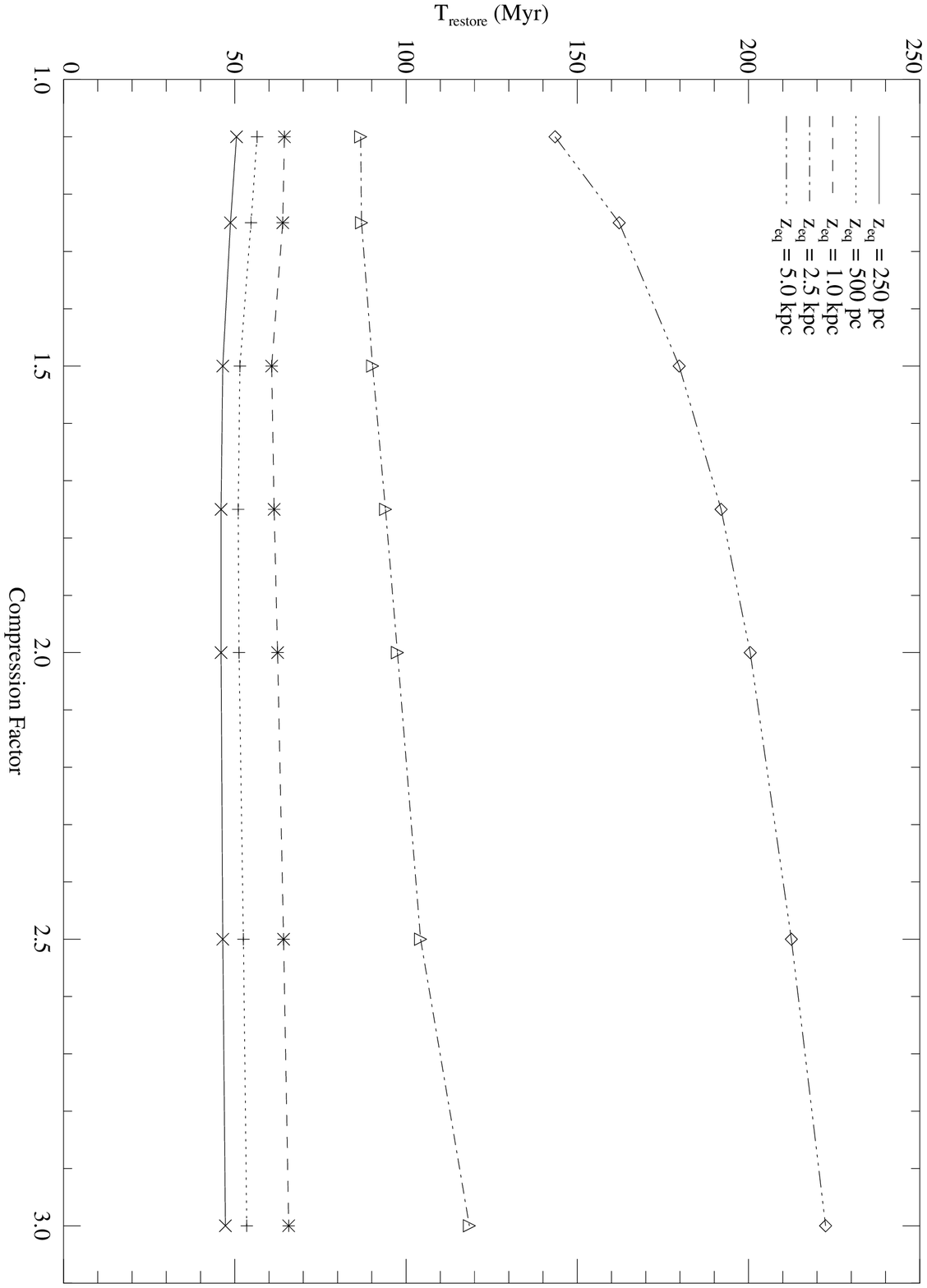]{Bounce time vs. compression factor for tracer
parcels with various initial heights. The atmospheric boundary is at 5 kpc.
\label{fig14}}

\figcaption[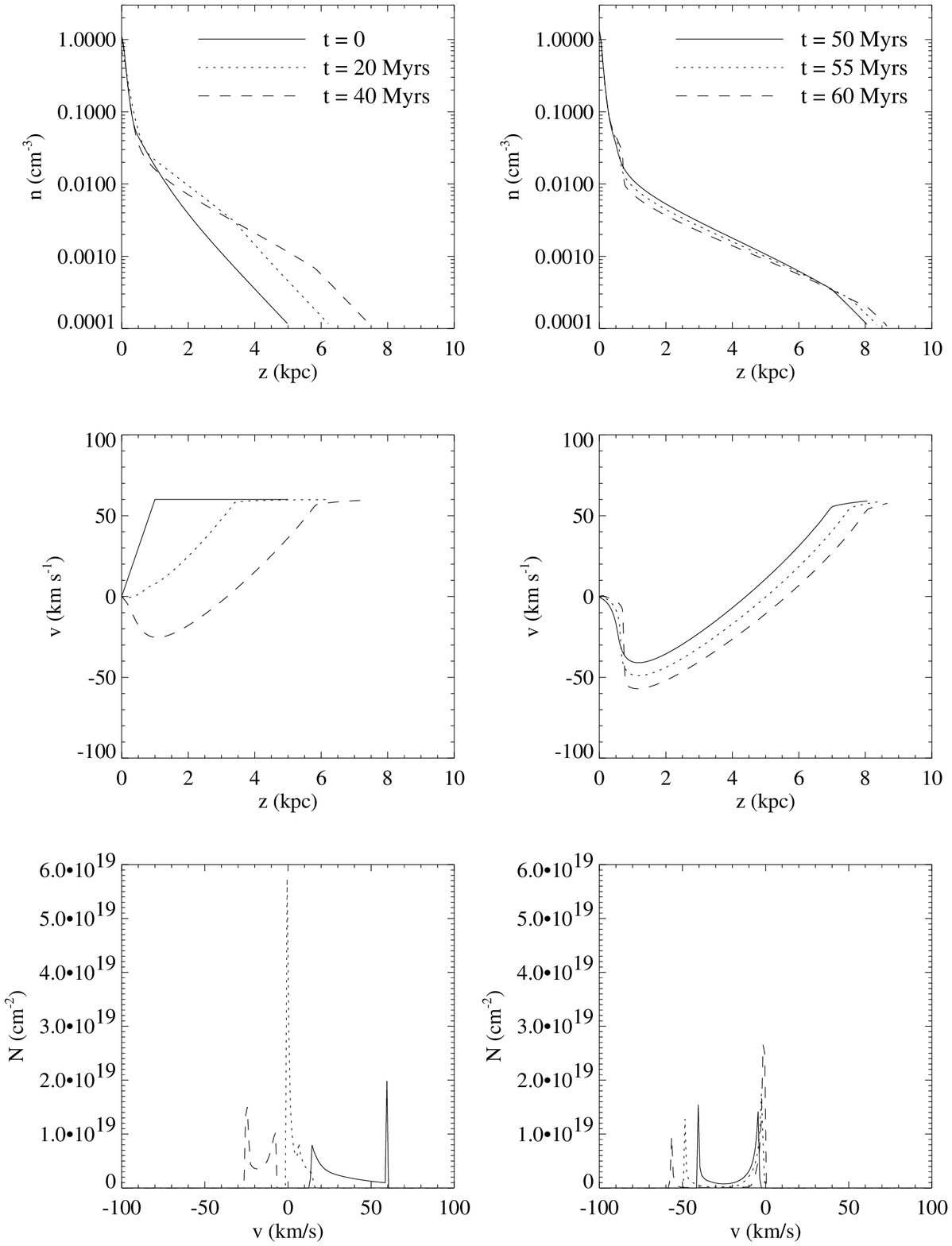]{Density distributions, velocity distributions,
and velocity spectra for 6 different times in the ``falling sky'' problem.
The top left graph shows the density distributions at 0, 20 and 40 Myr. The
middle left plot shows the velocity for these times. Note that the velocity
profile for $t = 0$ illustrates the form of the initial disturbance imposed 
on the atmosphere. The bottom left plot gives the column density in 1 km 
s$^{-1}$ bins as a function of velocity for these times. The right column
shows the same distributions at 50, 55, and 60 Myr.
\label{fig18}}

\clearpage

\figcaption[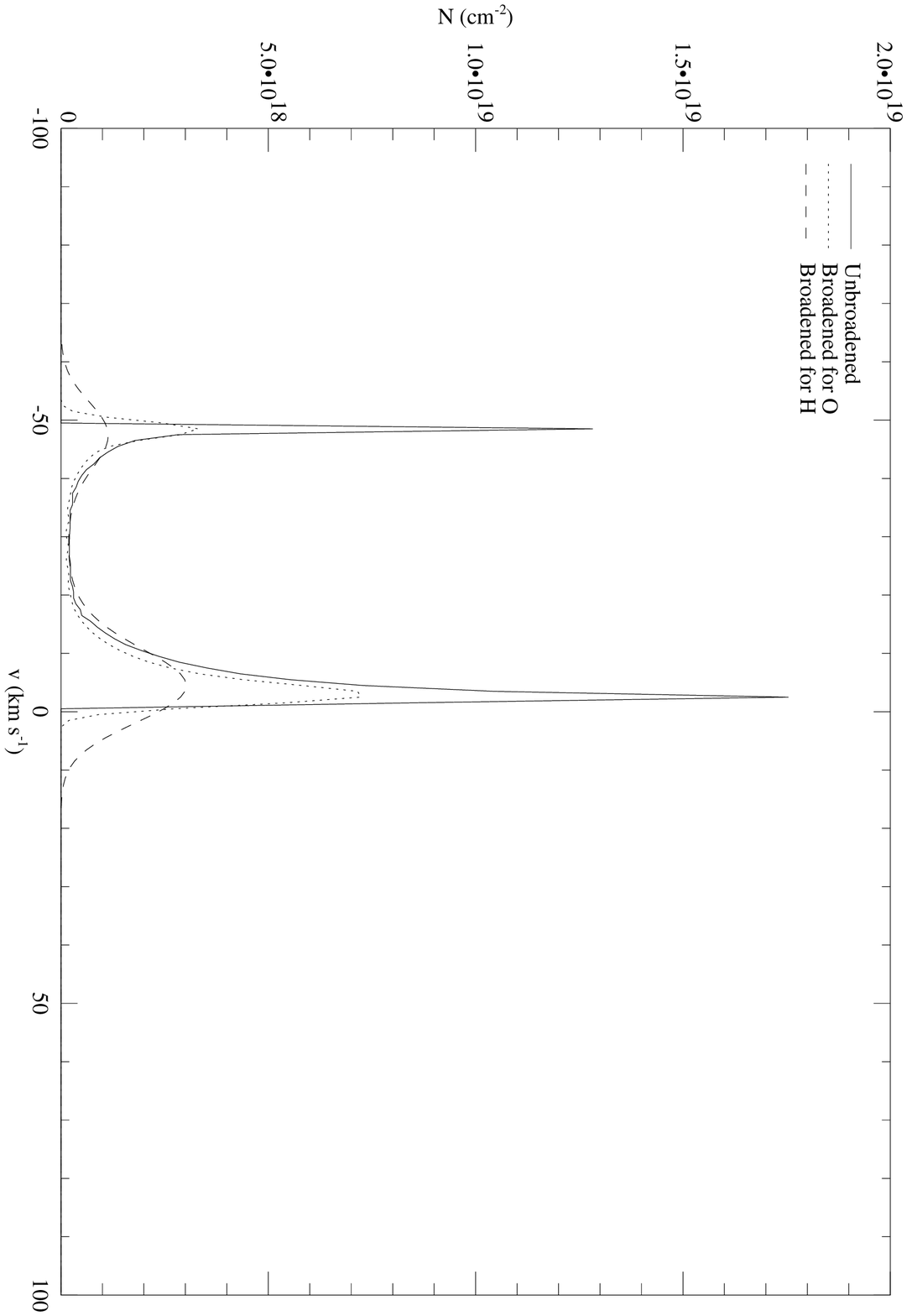]{Expanded view of the velocity spectrum at 55
Myr (with 1 km s$^{-1}$ bins in velocity space). The solid line shows the
spectrum with no broadening. The dotted and dashed lines show the spectrum
broadened by the thermal motions of oxygen and hydrogen atoms, respectively.
\label{fig19}}

\figcaption[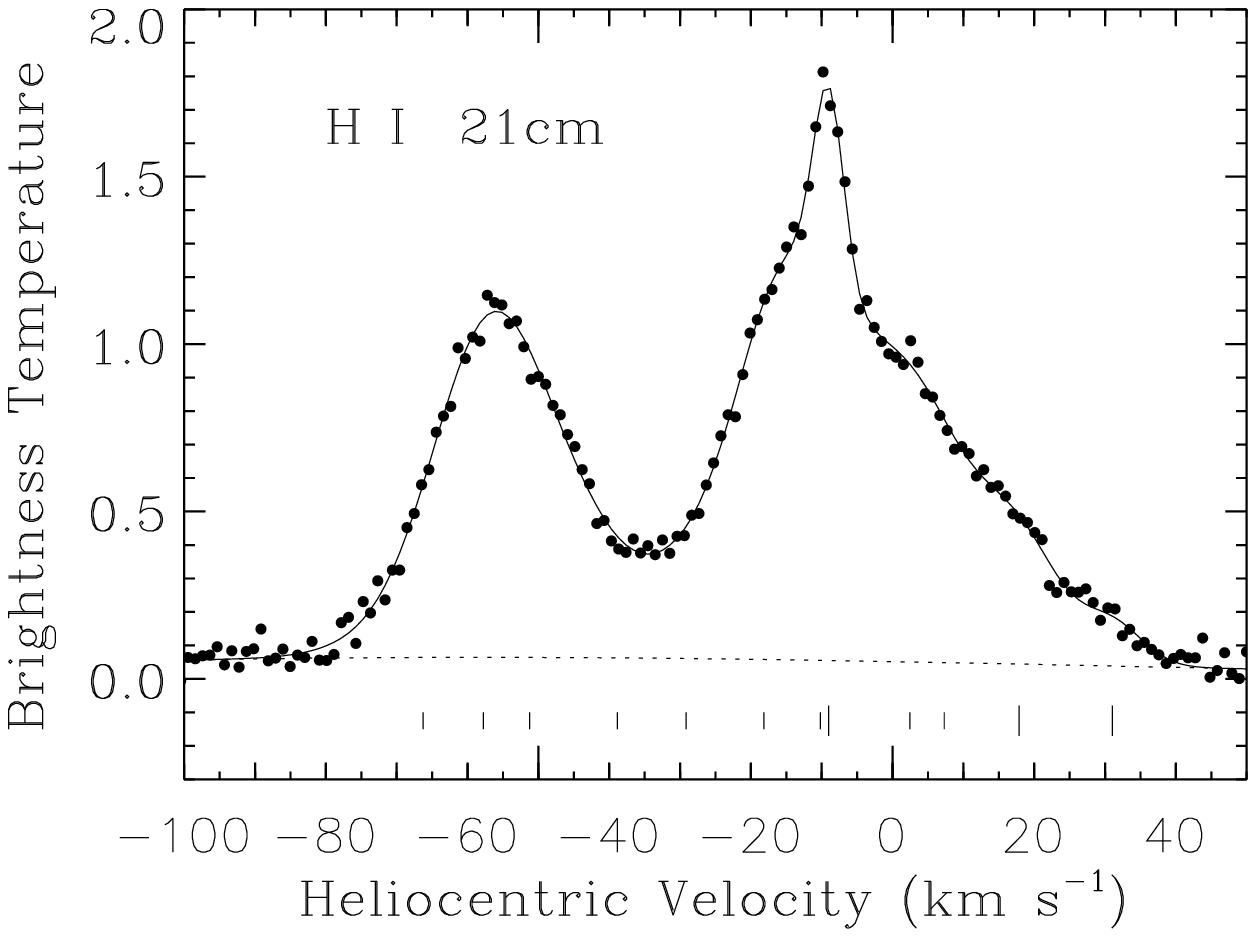]{Profile of 21 cm emissions toward HD 93521. This figure
was reproduced with permission from Figure 3 of Spitzer \& Fitzpatrick
(1993). The smaller tick marks below the dotted line indicate the velocity
components identified in the UV spectra. The larger tick marks are
additional components needed to fit the either or both of the Na I and H I
profiles.
\label{fig20}}

\figcaption[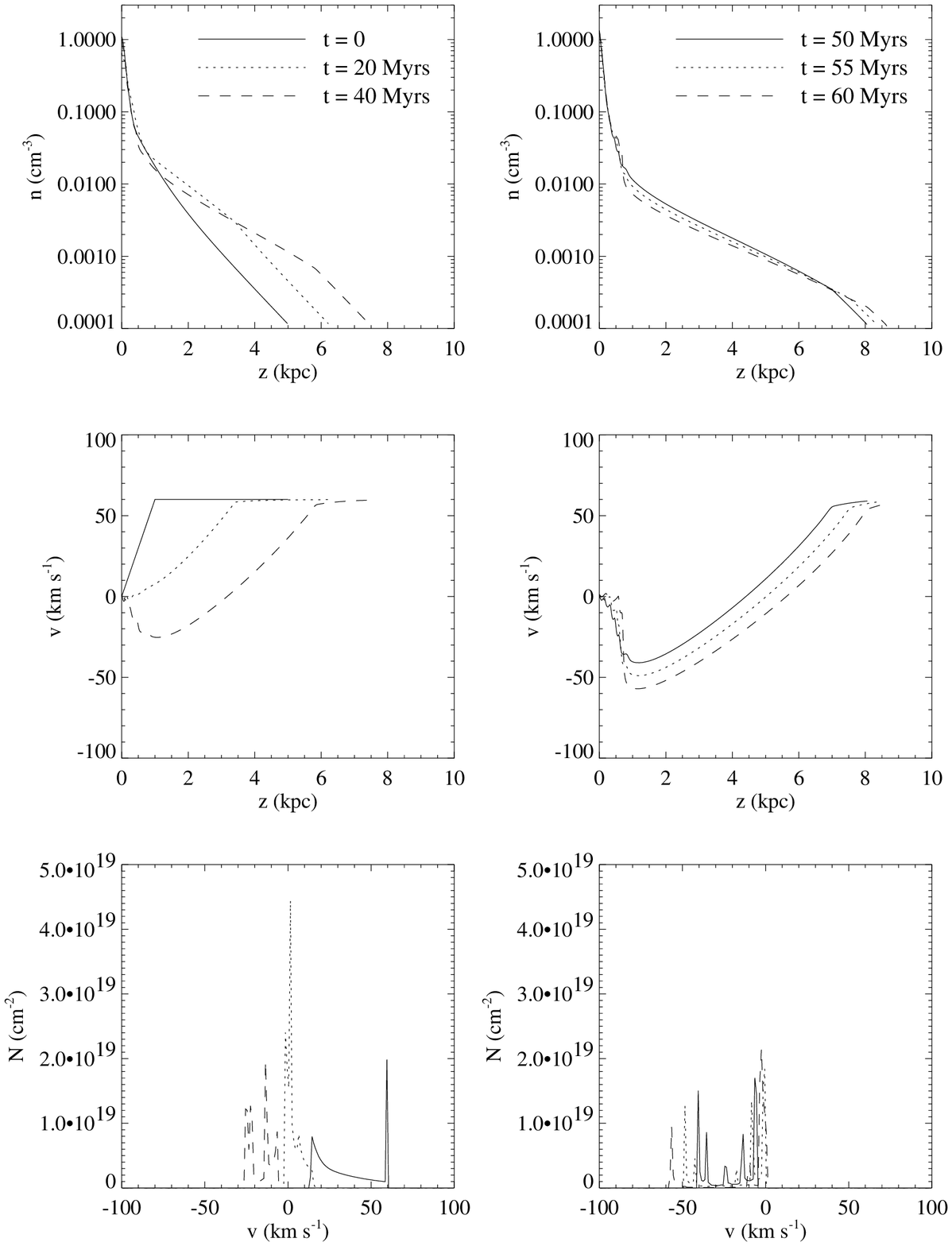]{Similar to Figure \ref{fig18}, but with small
waves, of midplane amplitude 5 km s$^{-1}$ and period 10 Myr, added. The left
column again shows plots for 0, 20, and 40 Myr; the right column shows times
of 50, 55, and 60 Myr. The small waves have little effect on the density (top)
and velocity (middle) profiles, but make a dramatic difference in the velocity
spectra (bottom).
\label{fig22}}

\figcaption[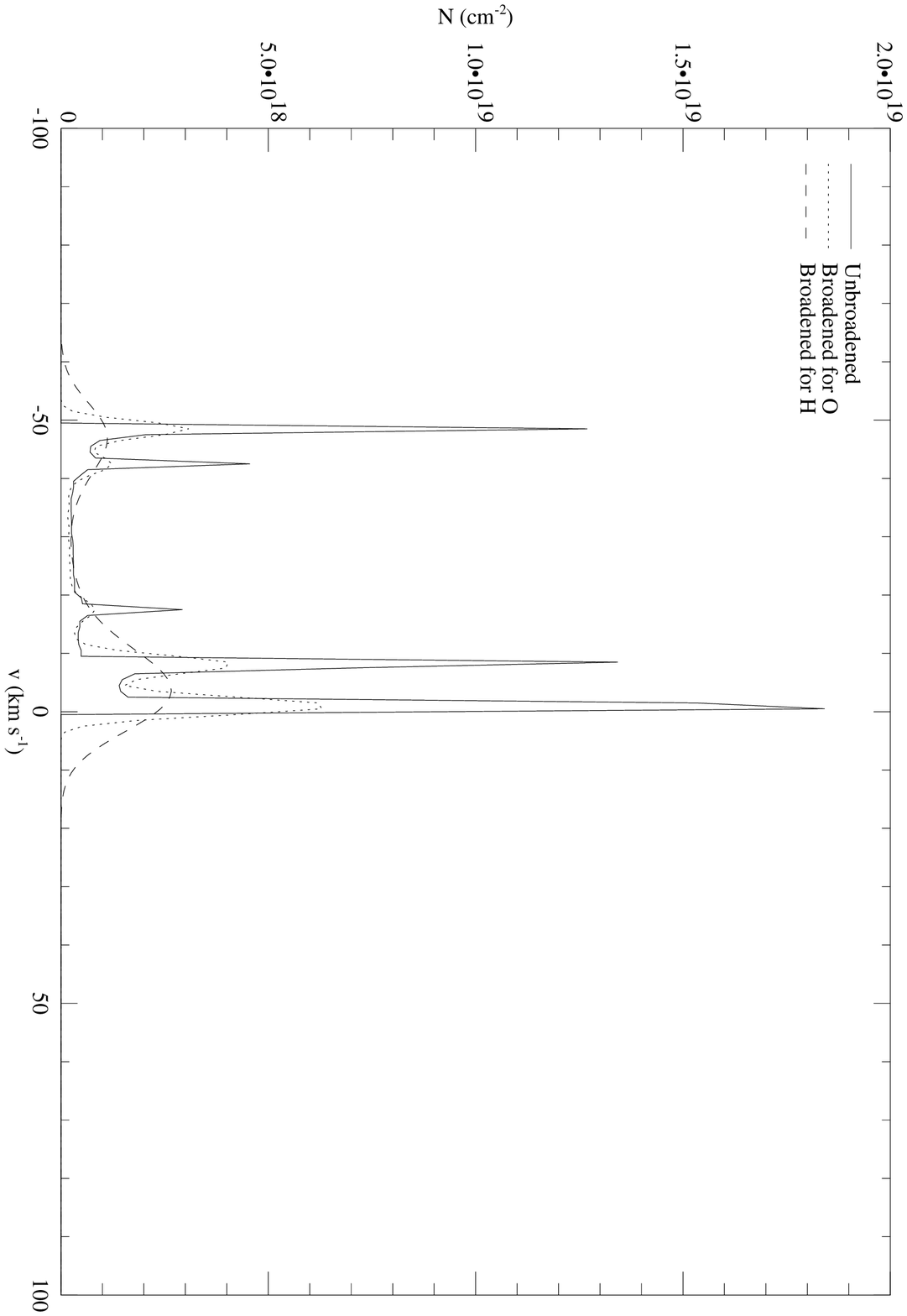]{Velocity spectrum at 55 Myr (in 1 km s$^{-1}$
bins) for the falling sky with small waves added. Also shown is the spectrum as
thermally broadened for O (dotted line) and H (dashed line).
\label{fig23}}

\figcaption[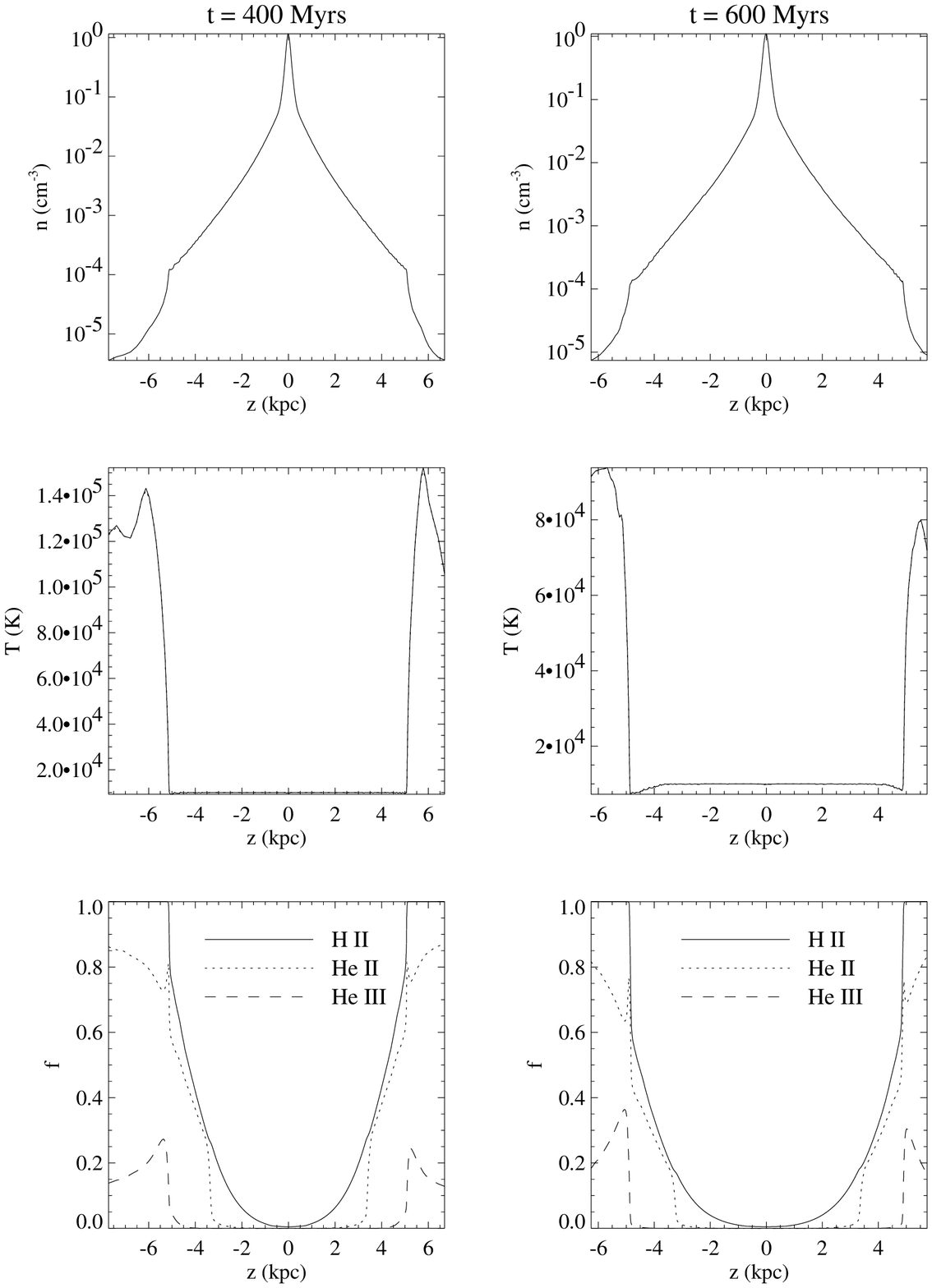]{The falling sky problem after 400 Myr (left) and 600 Myr (right),
showing the densities (top), temperatures (middle), and ion fractions
(bottom). Notice that hot, ionized, low-density layers have formed at the
outer edges of the atmosphere and remain there for a long time.
\label{fig24}}

\figcaption[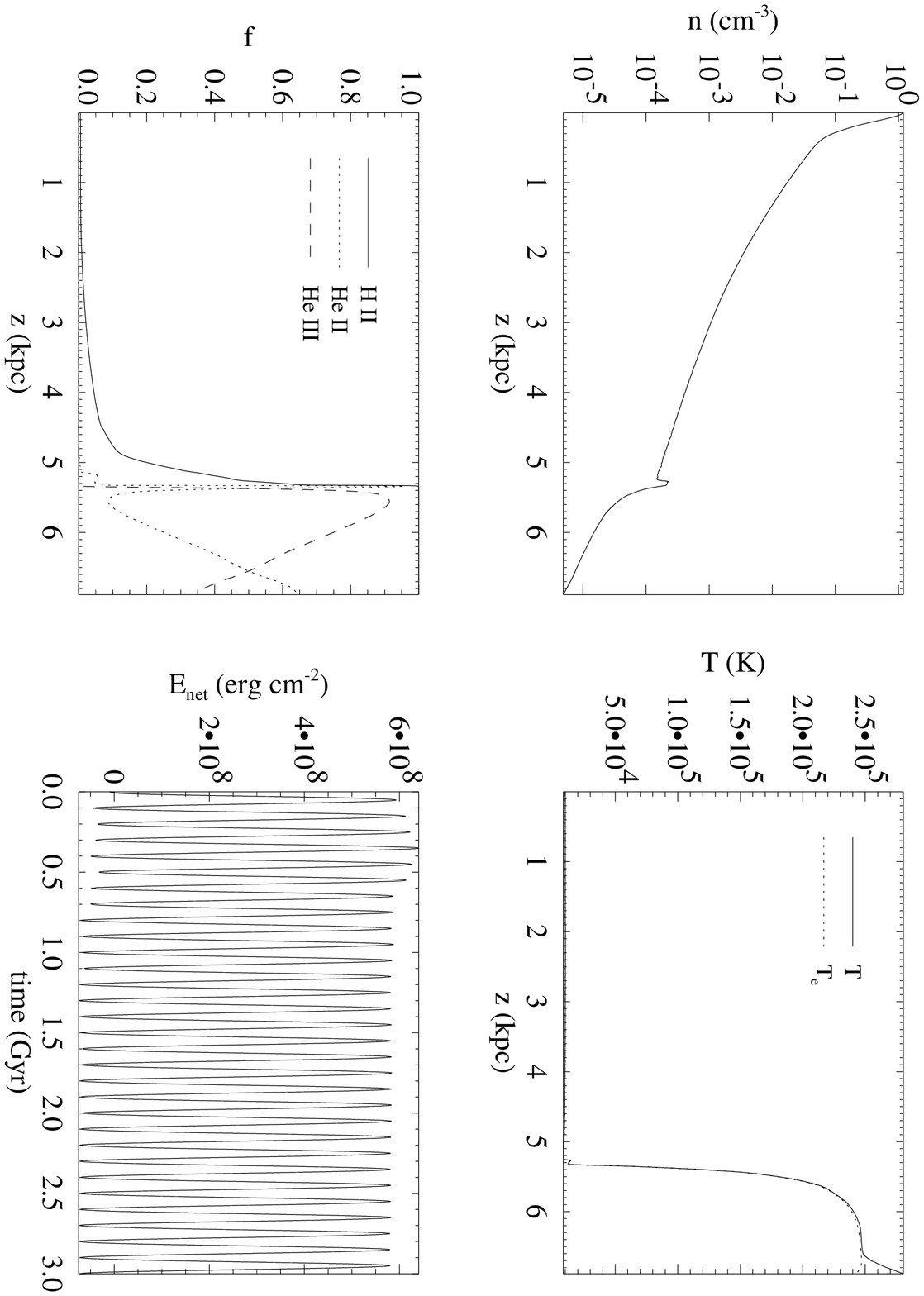]{Effects of a small periodic disturbance, of amplitude
2 km s$^{-1}$ and period 100 Myr, after 3 Gyr. Again a hot, ionized,
low-density layer has formed at the top of the atmosphere. The lower right
plot shows the change in total energy versus time. In spite of the 
oscillations associated with the disturbance, the average total energy does not change.
\label{fig25}}

\figcaption[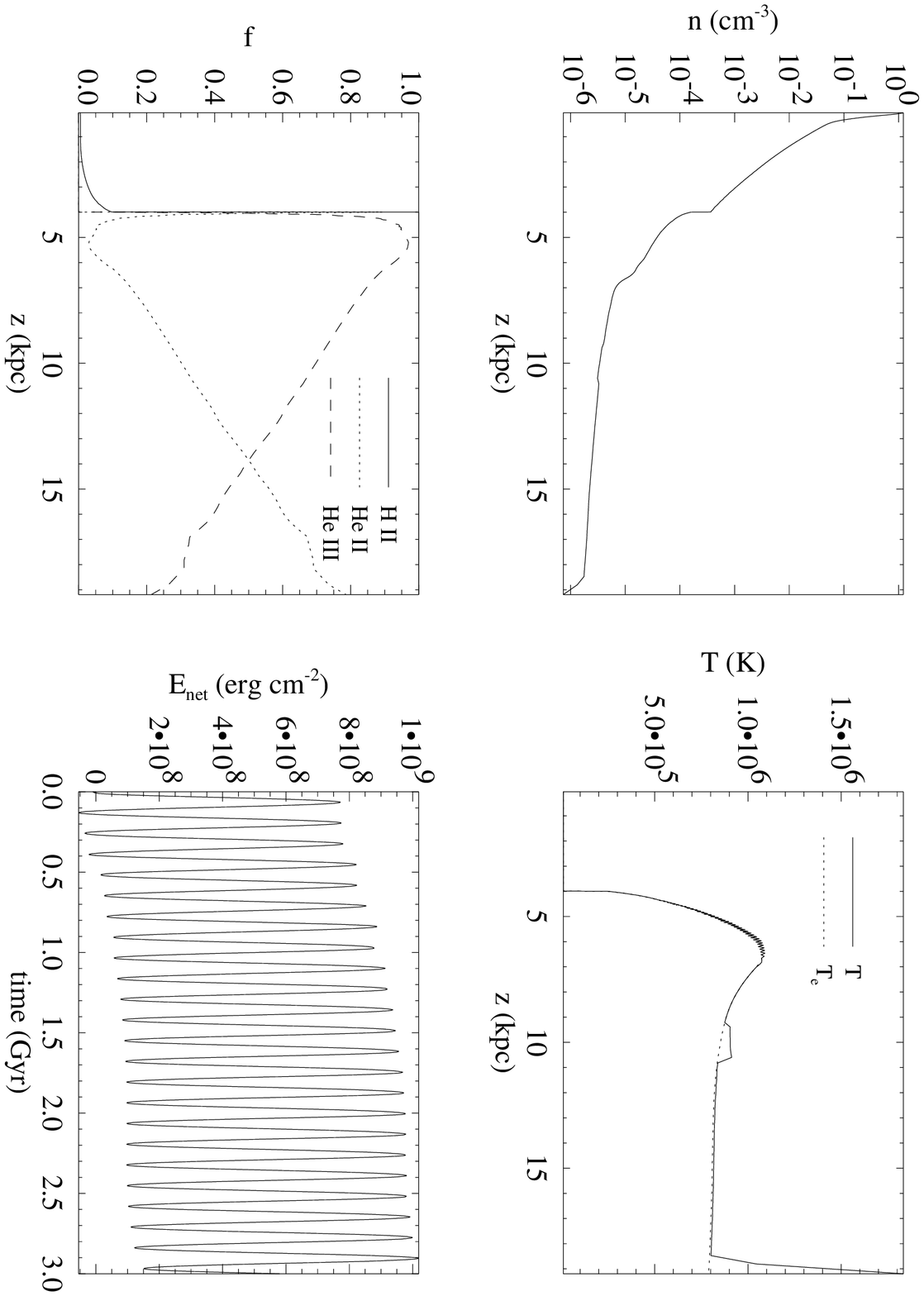]{Effects (after 3 Gyr) of a small periodic disturbance, 
of amplitude 2 km s$^{-1}$\ at midplane with
a period of 129 Myr, corresponding to the first symmetric mode of the
unperturbed atmosphere. The upper atmosphere is much hotter and more
distended than cases with periods of 100 and 133 Myr, and there is a
definite trend towards increasing energy.
\label{fig27}}

\clearpage

\figcaption[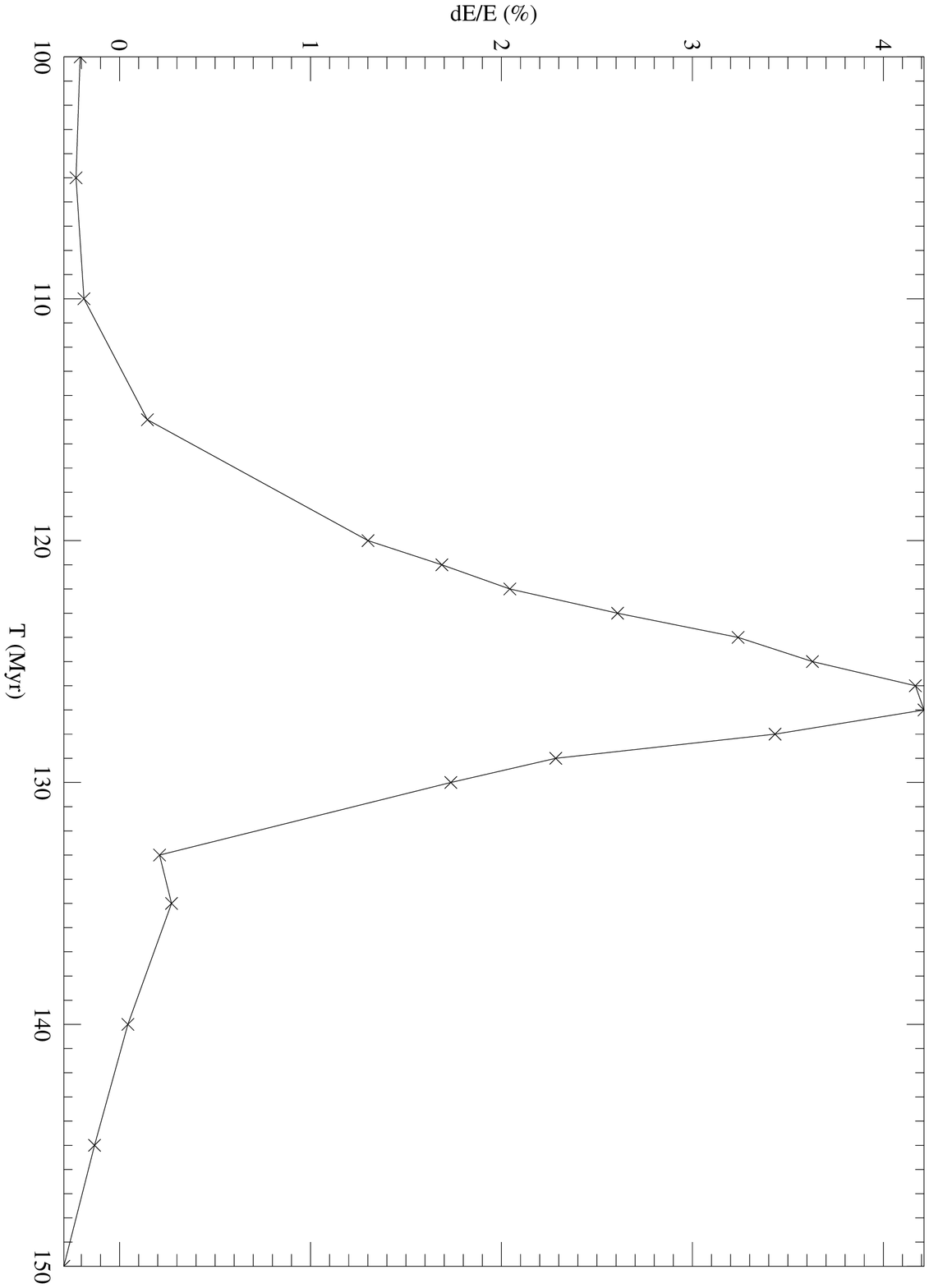]{Net change in total energy column density (as a percentage of
initial energy density) as a function of the period of the perturbation. The peak
response occurs at 127 Myr, which is slightly less than the resonant period 
of 129 Myr for the first symmetric mode of the unperturbed atmosphere.
\label{fig28}}

\figcaption[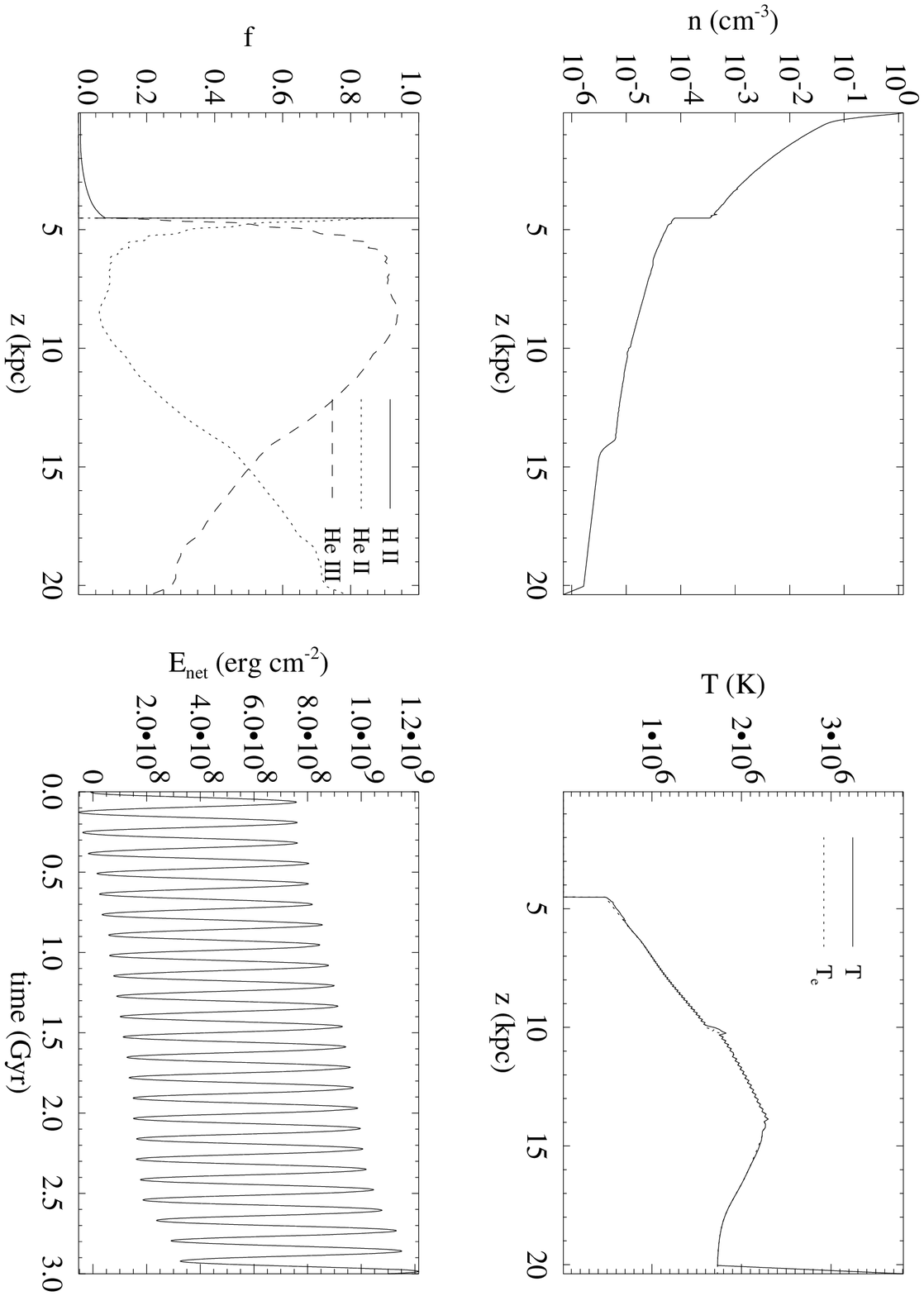]{Response of the atmosphere (after 3 Gyr) to
small waves at the peak period of 127 Myr. The waves have created a very hot,
very distended, and highly ionized layer at the top of the atmosphere. Also,
the energy continues to increase with no sign of leveling off.
\label{fig29}}

\end{document}